\documentclass{article}

\usepackage[preprint,nonatbib]{neurips_2020}

\usepackage[numbers]{natbib} 

\usepackage[utf8]{inputenc} 
\usepackage[T1]{fontenc}    
\usepackage{hyperref}       
\usepackage{url}            
\usepackage{booktabs}       
\usepackage{amsfonts}       
\usepackage{nicefrac}       
\usepackage{microtype}      

\usepackage{graphicx}
\usepackage{subcaption}
\usepackage{booktabs} 

\usepackage{amsmath}
\usepackage{amssymb}
\usepackage{mathtools}
\usepackage{amsthm}
\usepackage{bbm}

\newcommand{\indep}{\perp \!\!\! \perp}

\title{Generalized Permutation Framework for Testing Model Variable Significance}

\author{%
  Yue Wu \\
  \texttt{yueswu@uw.edu} \\
  \And
  Kenji Nakamichi \\
  \AND
  Russell Van Gelder \\
  \And
  Aaron Lee \\
}

\begin{document}

\maketitle

\begin{abstract}
A common problem in machine learning is determining if a variable significantly contributes to a model's prediction performance.
This problem is aggravated for datasets, such as gene expression datasets, that suffer the worst case of dimensionality: 
a low number of observations along with a high number of possible explanatory variables.
In such scenarios, traditional methods for testing variable statistical significance or constructing variable confidence intervals may not apply.
To address these problems, we developed a novel generalized permutation framework (GPF) for testing the significance of variables in supervised models.
Our permutation framework has three main advantages. 
First, it is non-parametric and does not rely on distributional assumptions or asymptotic results. 
Second, it is model agnostic and allows one to construct the null distribution for any metric of interest for a feature in a supervised model.
Third, it can overcome co-linearity in high dimensional datasets and determine contribution of each variable.
We demonstrate the performance of the generalized permutation framework on synthetic datasets versus existing methods,
and then applied it to multi-class classification of brain regions in RNA expression data,
and used this framework to show variable-level statistical significance and interactions. 
\end{abstract}

\section{Introduction}
The curse of dimensionality has often posed challenges to machine learning,
not only in terms of training a model, 
but also for understanding the trained model.
It is often as important to be able to interpret a fitted model.
For example, an interpretable model on gene-expression data should allow one to validate the model predictions externally using independent molecular assays and gain insight into diseases.
Furthermore, a model would be quantifiably interpretable if it identified the relative contributions of different genes to modeling and predicting disease,
thereby helping researchers prioritize the molecular assays.

Classically, the significance of variable contributions to models are measured by p-values.
For example, in generalized linear models the contribution of each explanatory variable can be quantified by their p-values or 95\% confidence intervals.
More generally, non-linear models, such as random forests and deep learning based models, have achieved success in modeling and predicting labeled targets in a supervised setting, 
but are less easily interpretable.
Frameworks have been proposed for interpreting these models, such as tree feature importance \citep{breiman2017classification}, gradient based attention and saliency maps \citep{selvaraju2017grad, simonyan2013deep} or game theoretic interpretations such as SHAP \citep{lundberg2017unified}.
One of the key shortcomings of these frameworks is the lack of traditional statistical measures for the explanatory variables.

In this paper, we present a novel generalized permutation framework (GPF) that provides traditional statistical measures such p-values and confidence intervals for explanatory variables in non-linear models. 
Our three main contributions are:
First, the GPF allows one to construct the null distribution for the contribution of any feature in a supervised model for any metric of interest.
Second, the GPF makes no distributional nor conditional assumptions about the explanatory variables or target variable.
Third, the GPF can overcome co-linearity problems in high dimensional datasets by applying permutations to subsamples of the variables.
The GPF was first demonstrated and compared to state-of-the-art models on synthetic datasets.
Then it was applied to a real-world RNA expression dataset, which suffer from the curse of dimensionality, 
where it was able to discover the genes that make significant contributions in classifying brain regions. 

\section{Review of related work}

\subsection{Model explainers}
In the case of complex models 
that allow convoluted interactions between variables, such as XGBoost \citep{chen2016xgboost} and Deep Neural Networks \citep{goodfellow2016deep},
there has been extensive research into frameworks that explain the model predictions.
Model explainers such as LIME \citep{ribeiro2016should} and SHAP \citep{lundberg2017unified} 
show the impact each observation $\{ x_{m j} \}_{j=1}^{P}$, where $P$ is the number of variables, in making prediction $\hat{y}_{m}$.
They can also measure the overall impact of each variable $\mathbf{x}_{\cdot, j}$ by combining the impact of each observation  $\{x_{i j}\}_{i=1}^{N}$.
\citet{kumar2020problems} showed that mathematical problems arise when SHAP is used for feature importance in tree models.
Furthermore, SHAP does not provide a confidence interval on the impact of each variable in the model.

\subsection{Randomization tests}
Model-X knockoffs and Conditional Randomization Test (CRT) were proposed together in \citep{candes2018panning}.
The CRT algorithm is reproduced in Algorithm~\ref{alg:CRT_algo}.
CRT's feature importance statistic $T_{j}(\mathbf{X}, \mathbf{y})$ was chosen to be the Lasso coefficient difference (LCD) statistic, as the authors focused on linear Gaussian models.
Moreover, CRT can be applied to non-linear models if the feature importance statistic is chosen to be model appropriate.
However, CRT suffers a major shortcoming,
as it requires knowledge of the conditional distribution to sample knockoffs (Algorithm~\ref{alg:CRT_algo} Eq.~\ref{eq:simConditional}).
The conditional and thereby marginal distributions of the variables, $\mathcal{F}(\mathbf{X})$,
is only known if the variables were chosen in designed experiments.
In general for most datasets, one needs to learn the joint distribution of the variables $\mathbf{X}$ and the response $\mathbf{y}$, as well as marginal distribution of the variables, $\mathcal{F}(\mathbf{X})$.
Therefore the $\mathcal{F}(\mathbf{X})$ requirement limits the usefulness of CRT.

Another area of concern was CRT's computational costs, as it involved re-training each of the $P \times R$ simulated design matrices against the response.
Model-X knockoffs was presented as a computationally optimized version of CRT for linear Gaussian models,
where all the knockoffs $\mathbf{X_{j}}^{(r)}$ are generated at once for all $j=1,...,P$,
and these knockoffs appended to the original $\mathbf{X}$ to create a new design matrix $\mathbf{X'}$ that is $N \times 2P$. 
Then Lasso is fit on $\mathbf{X'}$, with the intuition that the coefficients for the knockoffs $X_{j}^{(r)}$ as well as those variables $X_{j}$ that are independent of the response $\mathbf{y}$ will shrink to $0$.
However, the authors themselves showed that CRT had higher power than model-X knockoffs.

The Holdout Randomization Test (HRT) \citep{tansey2022holdout} is a specialized CRT that avoids re-training by splitting the dataset into training and test datasets. 
HRT fits the model on the training dataset, and evaluates on conditionally sampled test data, 
which again requires knowledge of the marginal distribution, $\mathcal{F}(\mathbf{X})$.
Philosophically, HRT is similar to SHAP as they measure the conditional mutual information between the variables and the response.

An alternate method is the Conditional Permutation Test (CPT) \citep{berrett2020conditional}, which uses permutations based on ordered statistics to perturb the $\mathbf{X_{p}}_{p=1,...,P}$ variables.
The ordered statistics still require approximations of the conditional distribution $\mathcal{F}(\mathbf{X_{j}} | \{ \mathbf{X_{p}} \}_{p \neq j})   $.

CRT, HRT and CPT all require some knowledge of the conditional distribution $\mathcal{F}(\mathbf{X_{j}} | \{ \mathbf{X_{p}} \}_{p \neq j})   $, which is not always available.
Furthermore, computation gains in HRT might indirectly obfuscate the ability of the statistical model used to overfit the training data.

%
\subsection{Permutation tests}
An alternative to conditional randomization, which requires conditional knowledge about the variables, is to use permutations.
A standard technique to construct the null distribution is by using the permutation plug-in estimate \citep{scheffe1943statistical, wald1944statistical, hoeffding1952large, fisher1949design, good2013permutation}.
In the permutation plug-in technique,
the null distribution for a statistic of interest is constructed by randomly permuting the labels, and then computing the statistic for the permuted label data and unpermuted covariates.
The intuition is that if the labels are scrambled, 
the relationship between the labels and the variable will be broken,
and no variable should be able to reliably predict the scrambled labels.
Thus the statistic computed on the scrambled labels is a distribution of the statistic when the null hypothesis is true. 

\citet{li2013finding} applied the permutation plug-in estimate to construct the null distributions of the resampled rank statistic of genes for classifying genetic conditions.
Their algorithm, SAMseq, is summarized in Algorithm~\ref{alg:permPlugin}. 
Genes that are differentially expressed can then be tested against their null distributions in a non-parametric way without relying on distributional assumptions as in the popular DESeq2 method \citep{love2014moderated}.
Label permutation has also been applied to evaluate model predictive performance in \citep{golland2003permutation, williamson2017nonparametric, altmann2010permutation}, and to
test model variable selection techniques on chemical compound data in \citep{lindgren1996model}.
Additionally, \citet{strobl2008conditional} permuted the variables in synthetic data to study how correlated variables affected the construction of trees in Random Forests. 
In contrast, we use permutations in a model-independent fashion to analyze statistical contributions of each variable. 

\section{Generalized Permutation Framework (GPF)}
In this paper, we extend the permutation plug-in estimate to construct null distributions for supervised models generally, and are not constrained to specific model types.
To do this, we propose the construction of two types of null distributions by using permutations on different inputs of the supervised model.
Let $\Phi(\cdot)$ be a permutation function that randomly permutes an input vector.
The two types of permutations can be expressed as:
\begin{enumerate}
    \item Permute labels $\mathbf{y}$: $\Phi_{y}(\cdot)=\Phi(\mathbf{y})$ \label{itm:permY}. 
    \item Permute a variable  $\mathbf{x}_{\cdot, j}=\{x_{ij}\}_{i=1}^{N}$, but leave $\mathbf{y}$ and $\{ \mathbf{x}_{\cdot, k}\}_{k \neq j}$ unchanged: $\Phi_{x}(\cdot) = \Phi(\mathbf{x}_{\cdot, j})$ \label{itm:permX}.
\end{enumerate}

\subsection{Subset-GPF}
A naive implementation of the generalized permutation framework (Naive-GPF) incorporates these permutations and is shown in Algorithm~\ref{alg:naiveGPF}.
Naive-GPF is as computationally as expensive as CRT.
Therefore we take a subset approach and introduce Subset-GPF in Algorithm~\ref{alg:subsetGPF}.
\begin{algorithm}[ht]
   \caption{Subset-GPF}
   \label{alg:subsetGPF}
\belowdisplayskip=-5pt 
\begin{algorithmic}[1]
   \STATE {\bfseries Input:} labels $\mathbf{y}=\{y_{i}\}_{i=1}^{N}$,  covariates $\mathbf{X}= \{ x_{ij} \}_{i=1,j=1}^{N,P}$, and supervised model $f$
   \STATE {\bfseries hyperparameters:} subsample size $K$, $M$ variable indices of interest, $\{j_{m} \in \{1, ..., P\}\}_{m=1}^{M}$ for the corresponding variables $\{ \mathbf{x}_{\cdot, j_{m}} \}_{m=1}^{M}$
   \FOR{$m=1$ {\bfseries to} $M$}
    \FOR{$r=1$ {\bfseries to} $R$}
       \STATE Sample $K-1$ times without replacement from $\{1,...,P \} \setminus j_{m}$ to get $\{ k' \}$
       \STATE Let $\mathbf{X}_{sub} = \{ \mathbf{x}_{\cdot, k}\}_{k \in \{ \mathbf{k'} \} }$
       \STATE Concatenate $\mathbf{x}_{\cdot, j_{m}}$ with $\mathbf{X}_{sub}$ 
       to obtain $\mathbf{Z}$ with dimensions $N \times K$
       \STATE Split $\mathbf{y}$ and $\mathbf{Z}$ into training and test sets. 
       \STATE Train and compute the test statistic $T^{(r)}_{j_{m}}$ from $\mathbf{y}$,  $\mathbf{Z}$ and $f$.
		\STATE Permute the variable of interest, $\mathbf{x'_{\cdot, j_{m}}} = \Phi_{x} (\mathbf{x}_{\cdot, j_{m}})$
       \STATE Concatenate $\mathbf{x'}_{\cdot, m}$ with $\mathbf{X}_{sub}$ to obtain $\mathbf{Z'}$
       \STATE Train and compute test statistic $T'^{(r)}_{j_{m}}$ from $\mathbf{y}$, $\mathbf{Z'}$ and $f$ using same train/test split. 
    \ENDFOR
   \ENDFOR
   \STATE {\bfseries Output}: \{$\{ T'^{(r)}_{j_{m}} \}_{r=1}^{R} \}_{m=1}^{M}$ and $\{ \{ T^{(r)}_{j_{m}} \}_{r=1}^{R} \}_{m=1}^{M}$ 
\end{algorithmic}
\end{algorithm}
The intuition of Subset-GPF is to reduce the computation costs of re-training experienced in CRT and Naive-GPF, by re-training on subsets $\mathbf{Z}$ that are $N \times K$, instead of on the full design matrix $\mathbf{X}, N \times P$.
Subset-GPF not only has computational advantages, but can avoid potential co-linearity problems, 
especially in datasets with $N \ll P$.
The subsamples can mitigate confounding model contributions between highly correlated variables,
by reducing the co-selection of correlated variables.

The hyperparameter $K$ can be selected depending on the dataset and learning model.
For models that learn covariance structures, which require $K^{2}$ observations, then let $K=\left \lfloor{\sqrt{N} } \right \rfloor$.
Alternatively, $K$ can be chosen to balance model needs and model computation speeds. 
A full discussion of the choice for $K$ for various types of learning models, 
as well as the proof of the consistency of size $K$ Subset-GPF,
is provided in the Supplement.



\subsection{GPF vs existing methods} \label{GPF_vs_existing}
Table~\ref{table:frameworkComparison} summarizes the properties of the various frameworks in terms of 
datasets perturbed, 
the perturbation function, 
and if the model $f$ retrained after data perturbation.
The last two columns spotlight, the null hypothesis and the test statistics.
Note that for models that do not retrain, such as SHAP and HRT, the null is that a variable $X_{j}$ is independent of  $Y$ given the pre-trained $f$ and the other variables ${X_{p}}_{p \ne j}$.

\begin{table}[t]
\setlength\tabcolsep{3.75pt}
\vskip -0.15in
\caption{Framework Comparison}
\label{table:frameworkComparison}
\begin{small}
\begin{tabular}{cccccc}
\toprule
Framework & Dataset & Perturbation & Re-train $f$ & $H_{0}$ & Test Stat  \\
\midrule  
SHAP &  Train | Test & $\text{exclude}(X_{j})$ & No & $X_{j} \indep Y | f, \{ X_{p} \}_{p \ne j}$ & Model score \\
Perm. Importance & Train | Test & $\Phi(X_{j})$ & No & $X_{j} \indep Y | f, \{ X_{p} \}_{p \ne j}$ & Model score \\ 
HRT &  Test & $\mathcal{F}(X_{j} | \{ X_{p} \}_{p \ne j})$ & No & $X_{j} \indep Y | f, \{ X_{p} \}_{p \ne j}$ & Model score \\
HRT-CV &  Test & $\mathcal{F}(X_{j} | \{ X_{p} \}_{p \ne j})$ & CV & $X_{j} \indep Y | f, \{ X_{p} \}_{p \ne j}$ & Model score \\
CRT &  Train & $\mathcal{F}(X_{j} | \{ X_{p} \}_{p \ne j})$ & Yes & $X_{j} \indep Y | \{ X_{p} \}_{p \ne j}$ & LCD \\
GPF &  Train | Test & $\Phi(X_{j})$ & Yes & $X_{j} \indep Y | \{ X_{p} \}_{p \ne j}$ & LCD, Model score \\
Subset-GPF & Train | Test & $\Phi(X_{j})$ & Yes & $X_{j} \indep Y$ & LCD, Model score \\
\bottomrule
\end{tabular}
\end{small}
\vskip -0.15in
\end{table}

\section{Experiments}
We first compare Subset-GPF (Algorithm~\ref{alg:subsetGPF}) on synthetic datasets versus CRT and HRT, 
in terms of 
true positive rate (TPR), 
the false discovery rate (FDR) and 
the $F_{1}$ score.
Then we apply GPF to a publicly available RNA expression dataset from the Allen Institute study on aging brains \citep{miller2017neuropathological}.
The experiments were run a combination of local CPU servers and AWS EC2 CPU servers.


\subsection{Linear Gaussian data}
Let $\mathbf{X}$ be the design matrix of $N$ observations of $P$ variables, 
and $\mathbf{y}$ the corresponding response vector.
The dimensions were chosen with $N=250$ and $P=400$ to represent a sparse dataset.
Moreover, let $S$ denote the set of real explanatory variables with its size $|S|=20$.
Then $\mathbf{X}$ and $\mathbf{y}$ are generated as follows:
\begin{align}
	\mathbf{y} &\sim \mathcal{N}(\beta \mathbf{X}, \sigma^{2}_{\mathbf{y}}) \\
	\mathbf{X} &\sim \mathcal{N}(0, \Sigma_{\mathbf{x}}), \quad \beta_{p} = 0 \quad \text{if} \; X_{p} \notin S \\
	\Sigma_{\mathbf{X}}^{ij} &= 
	\begin{cases}
	 \sigma_{x}^{2}, & \text{if}\ i=j \\
	 \rho, &\text{if}\ i \neq j 
	\end{cases} 
\end{align}
To compare the performance of Subset-GPF, CRT and HRT on different signal to noise datasets, 
the dispersion of the variables and response were fixed, 
with $\sigma_{x}^{2} =1$ and  $\sigma_{y}^{2} =1$,
Moreover, the correlation was fixed $\rho=0.3$,
while $\beta$, was varied from $[0.5, 0.75, 1, 1.5, 2, 2.5, 3, 5, 7.5, 10]$.
$5$ datasets were generated from this data setup,
and Subset-GPF, CRT and HRT were each run on each dataset using Lasso as the supervised learning model.
Subset-GPF was run according to Algorithm~\ref{alg:subsetGPF},
with $M=P$ for full coverage, 
$K=\left \lfloor{\sqrt{N} } \right \rfloor= 15$ 
and $R = 400$.

Similarly, to investigate the effect of correlated confounding variables, 
we let $\beta=1$, $\sigma_{x}^{2} =1$ and  $\sigma_{y}^{2} =1$, 
while $\rho$ was varied from $0$ to $0.9$ in $0.1$ increments.
$5$ datasets were generated from this data setup,
and Subset-GPF, CRT and HRT were run on each dataset using Lasso as the training model.

\subsection{Binomial transformed linear Gaussian data}
The setup makes the following modification with the response $\mathbf{y}$ now binary and are samples from the corresponding binomial distribution:
\begin{align}
	\mathbf{y} &\sim \mathcal{\text{Bin}}(1, \beta \mathbf{X})  \label{eq:y_logistic}
\end{align}
Just as in the linear Gaussian setup, 
$5$ datasets with varying $\beta$ were generated with all other parameters fixed. 
Then Subset-GPF and CRT run on each dataset now using Logistic Regression with a fixed L1 penalty as the training model.
Next to investigate the confounding effect of correlated variables in this setup, $\beta$ was fixed and $\rho$ varied, and Subset-GPF and CRT run on the datasets generated from this condition.

\subsection{Step-wise data}
To investigate the performance of Subset-GPF on nonlinear models, we generated the response according to a step function:
\begin{align}	
	\mathbf{z} \sim &\mathcal{N}(\beta \mathbf{X}, 1) \\ \nonumber
		y_{i} &= 
		\begin{cases}
			1, & \text{if} \quad z_{i} \in (\mathbf{z}_{25\%}, \mathbf{z}_{75\%} ) \\	
			0, & \text{otherwise} 														 
		\end{cases} 
\end{align}
The step-wise data can represent biological activations that are on ($y_{i}=1$), 
when conditions are on in the middle of the range and off when conditions are in the tails.
Subset-GPF and CRT were run on $5$ datasets each for varying $\beta$ and $\rho$ for this step-wise setup,
with the training model being XGBoost, a gradient boosted tree model \citep{chen2016xgboost}.

\subsection{Allen brain data experiments}
The Allen Institute dataset on aging brains \citep{miller2017neuropathological} consisted of $337$ samples from $107$ brains.
The samples were collected from four areas in the brain: a) parietal cortex, b) temporal cortex, c) frontal white matter, and d) hippocampus. We used this publicly available gene-expression data with normalized fragments per kilobase of transcript per million (fpkm), corrected for batch and RNA quality using the RSEM pipeline \citep{li2011rsem}.
The final number of genes in the normalized data was $P=50281$.
Consequently, the targets $\mathbf{y}$ had $N=337$ observations, each corresponding to one of four brain regions,
and the variable matrix $\mathbf{X}$ had dimensions $337 \times 50281$.
For the Allen brain data, Subset-GPF was applied to identify the genes that are significantly predictive of brain region under the XGBoost model.

\section{Results}
\subsection{Synthetic data results}
 
\begin{figure}[htb]\centering
\begin{subfigure}{.33\linewidth}
\includegraphics[width=1.2\linewidth]{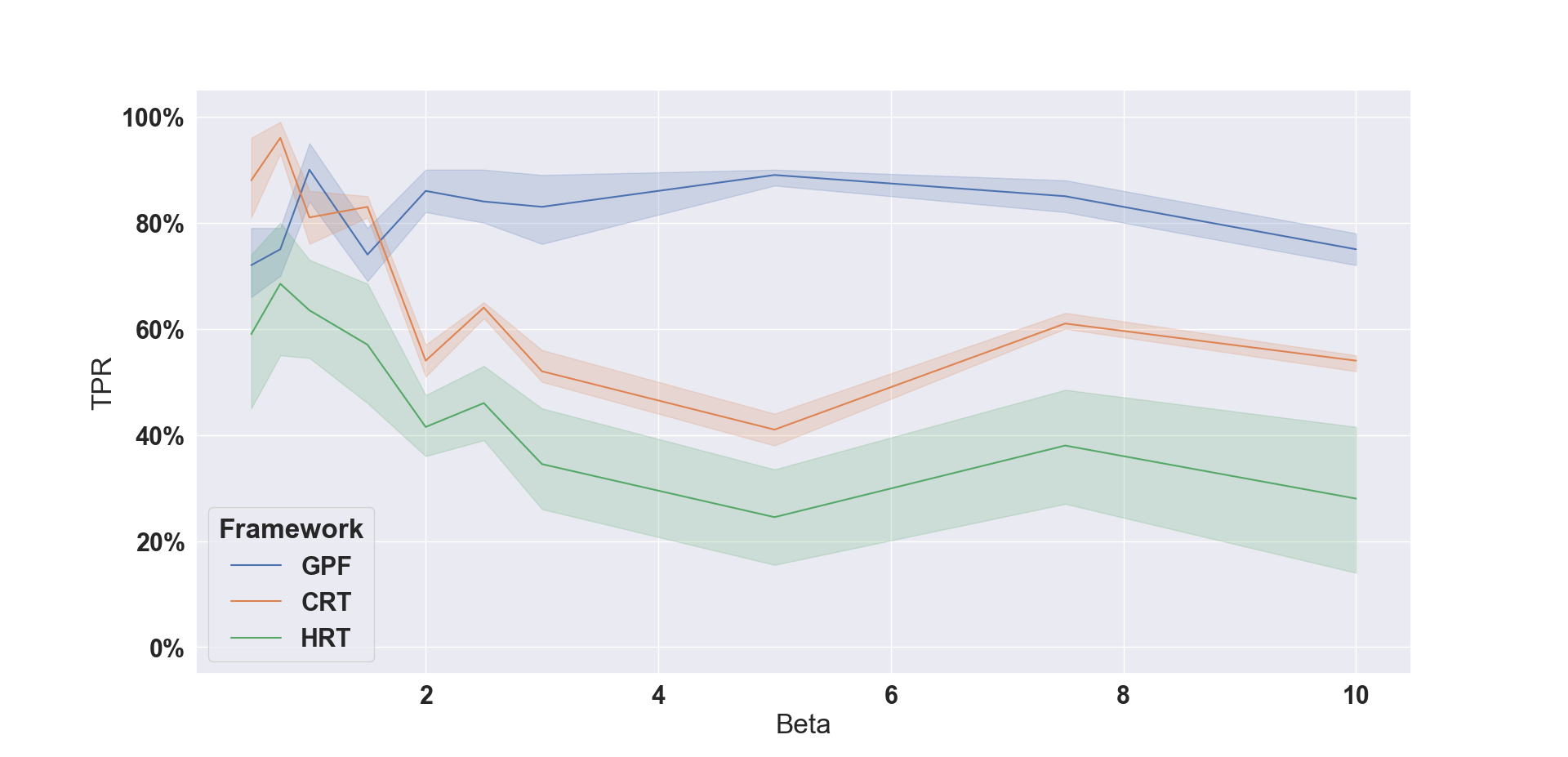}
\caption{TPR vs $\beta$}\label{fig.a}
\end{subfigure}%
%
%
\begin{subfigure}{.33\linewidth}
\includegraphics[width=1.2\linewidth]{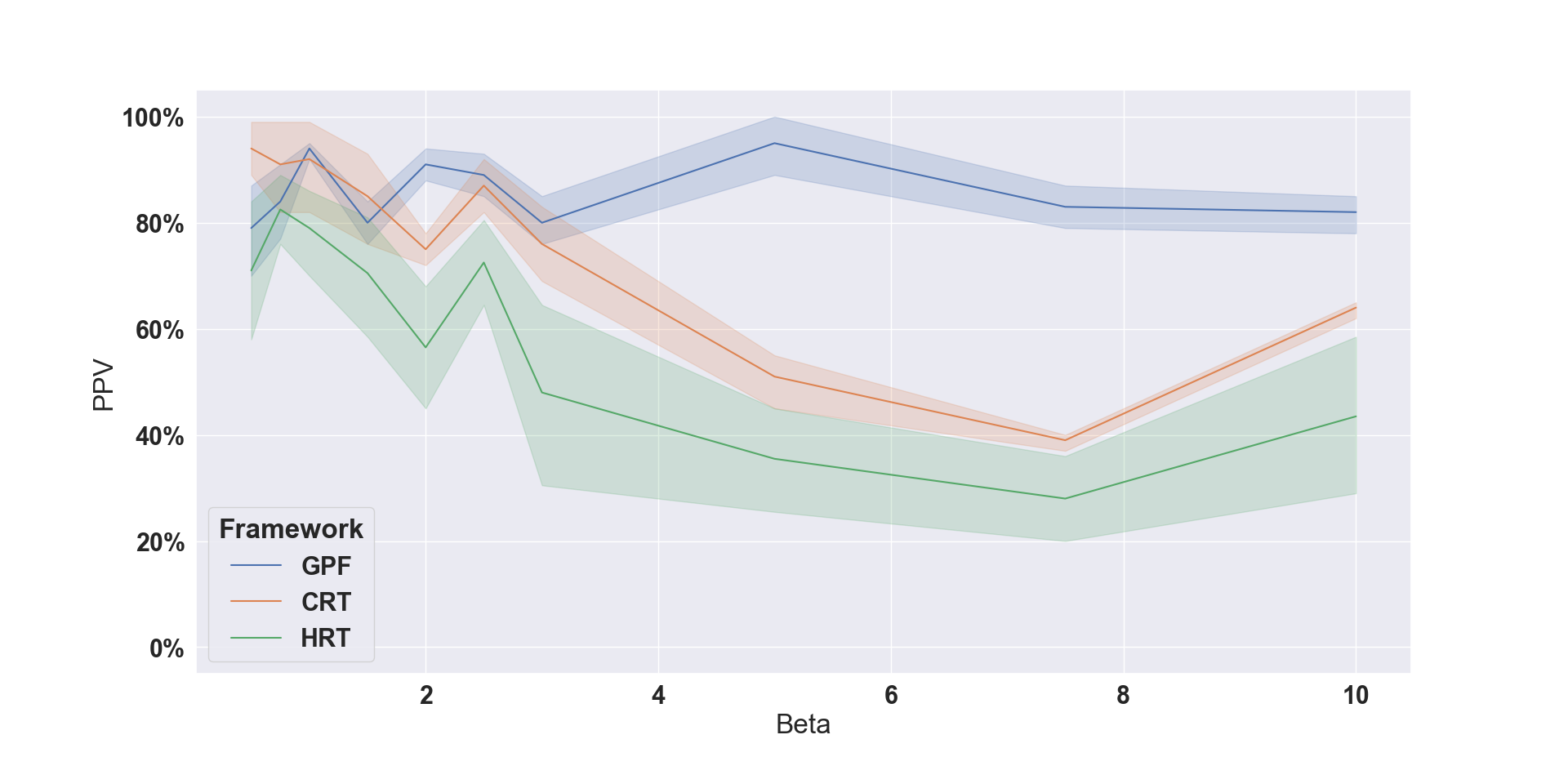}
\caption{PPV vs $\beta$}\label{fig.b}
\end{subfigure}
\begin{subfigure}{.33\linewidth}
\includegraphics[width=1.2\linewidth]{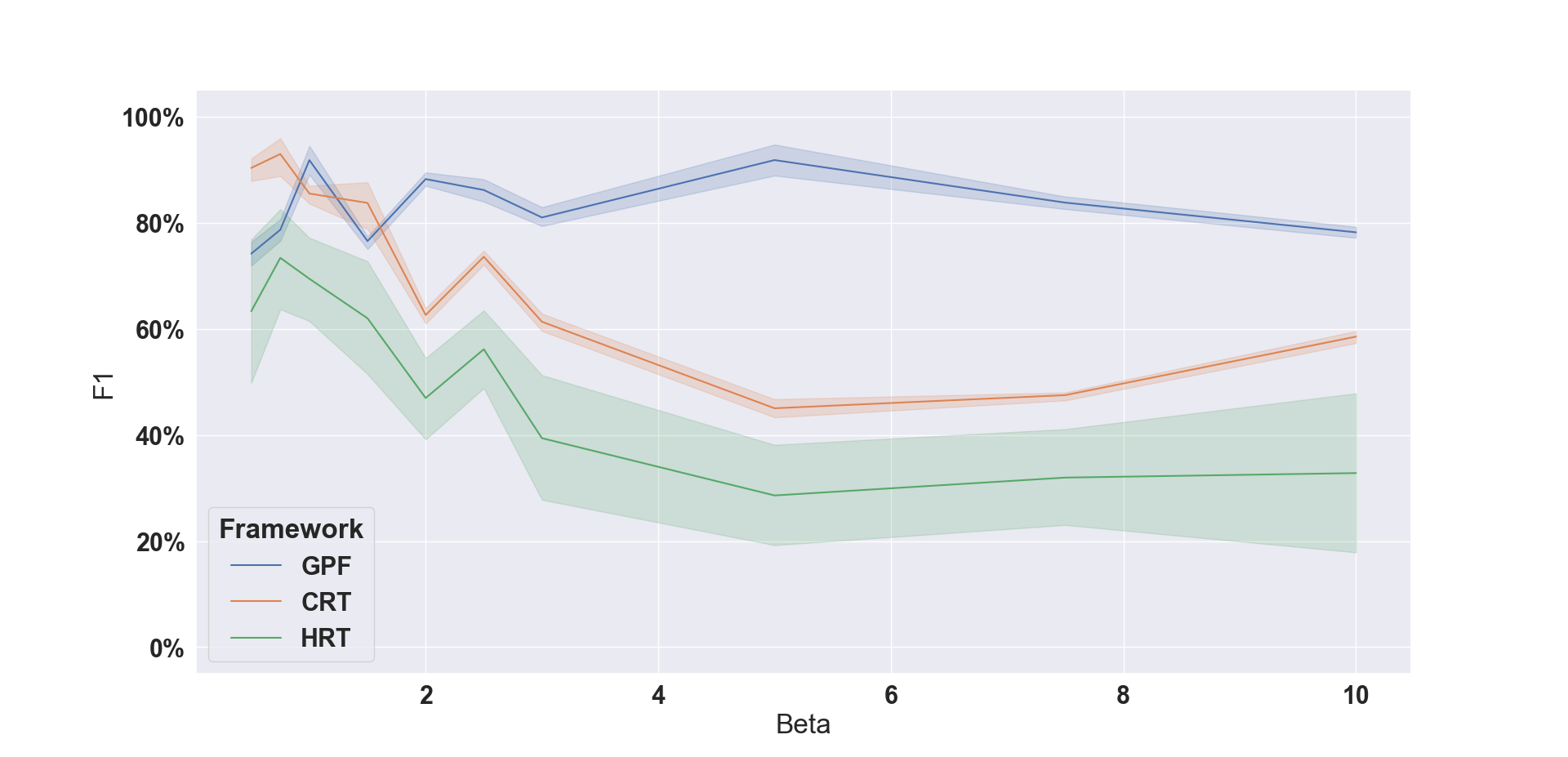}
\caption{$F_{1}$ vs $\beta$}\label{fig.c}
\end{subfigure}
\\
\begin{subfigure}{.33\linewidth}
\includegraphics[width=1.2\linewidth]{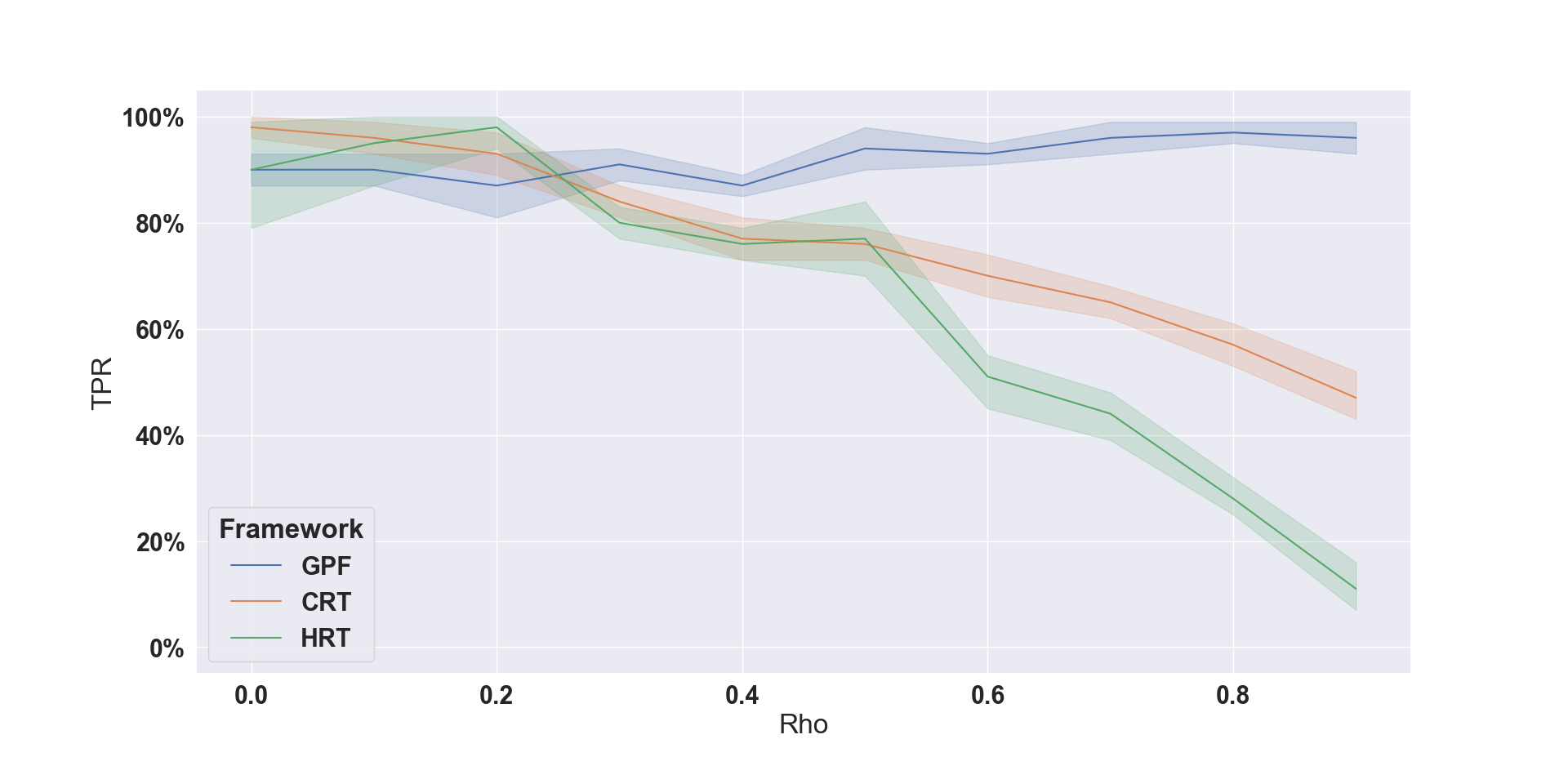}
\caption{TPR vs $\rho$}\label{fig.d}
\end{subfigure}
%
\begin{subfigure}{.33\linewidth}
\includegraphics[width=1.2\linewidth]{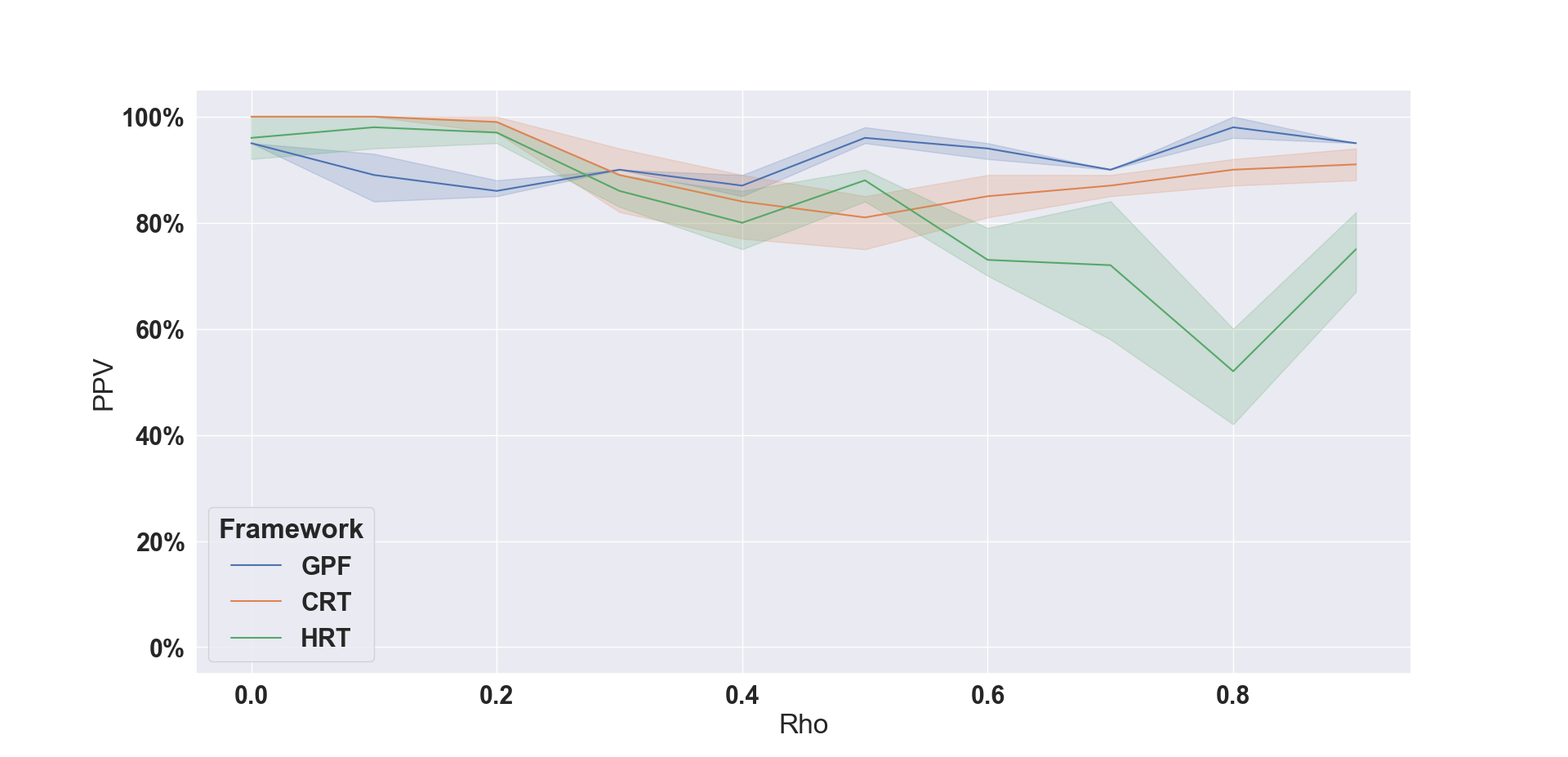}
\caption{PPV vs $\rho$}\label{fig.e}
\end{subfigure}%
\begin{subfigure}{.33\linewidth}
\includegraphics[width=1.2\linewidth]{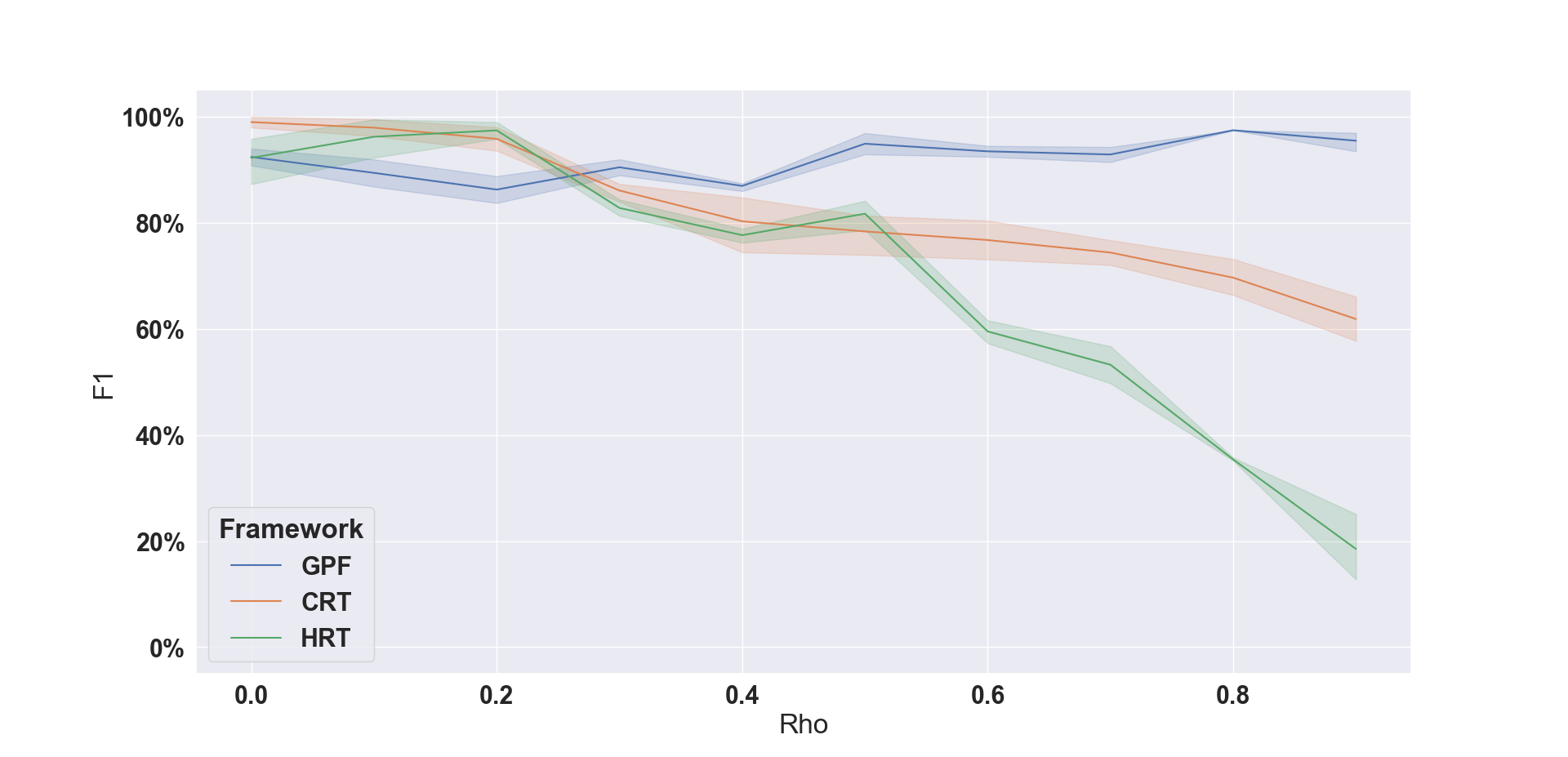}
\caption{$F_{1}$ vs $\rho$}\label{fig.f}
\end{subfigure}%
\caption{TPR, PPV and $F_{1}$ for GPF (blue), CRT (orange) and HRT (green) against signal strengths $\beta$ in the top row, and against variable correlation $\rho$ on bottom for the linear Gaussian data.}\label{fig:lassoComp}
\vskip -0.15in
\end{figure}

The Lasso coefficient difference (LCD) was the statistic used to determine which explanatory variables were not independent of the response for the linear Gaussian generated data for CRT and GPF,
while HRT used the test MSE as its empirical risk.
The TPR, positive predictive value (PPV = 1 - FDR), and $F_{1}$ are shown in Figure~\ref{fig:lassoComp}.
The solid lines shows the mean, while the shaded areas correspond to the 95\% CI determined by the 5 replication experiments.

When $\beta$ is small, all frameworks miss some of the true explanatory variables, 
and TPR increases as $\beta$ increases.
In contrast, when the variable correlation $\rho$ increases above 0.6, 
CRT and HRT struggle to find all true explanatory variables, 
while Subset-GPF still identifies nearly all.
We note that this correlation setting is harder than in \citep{candes2018panning}, which only considered auto-correlation, and not correlation across the entire variable set.

\begin{figure}[htb]\centering
\begin{subfigure}{.50\linewidth}
\includegraphics[width=1.15\linewidth]{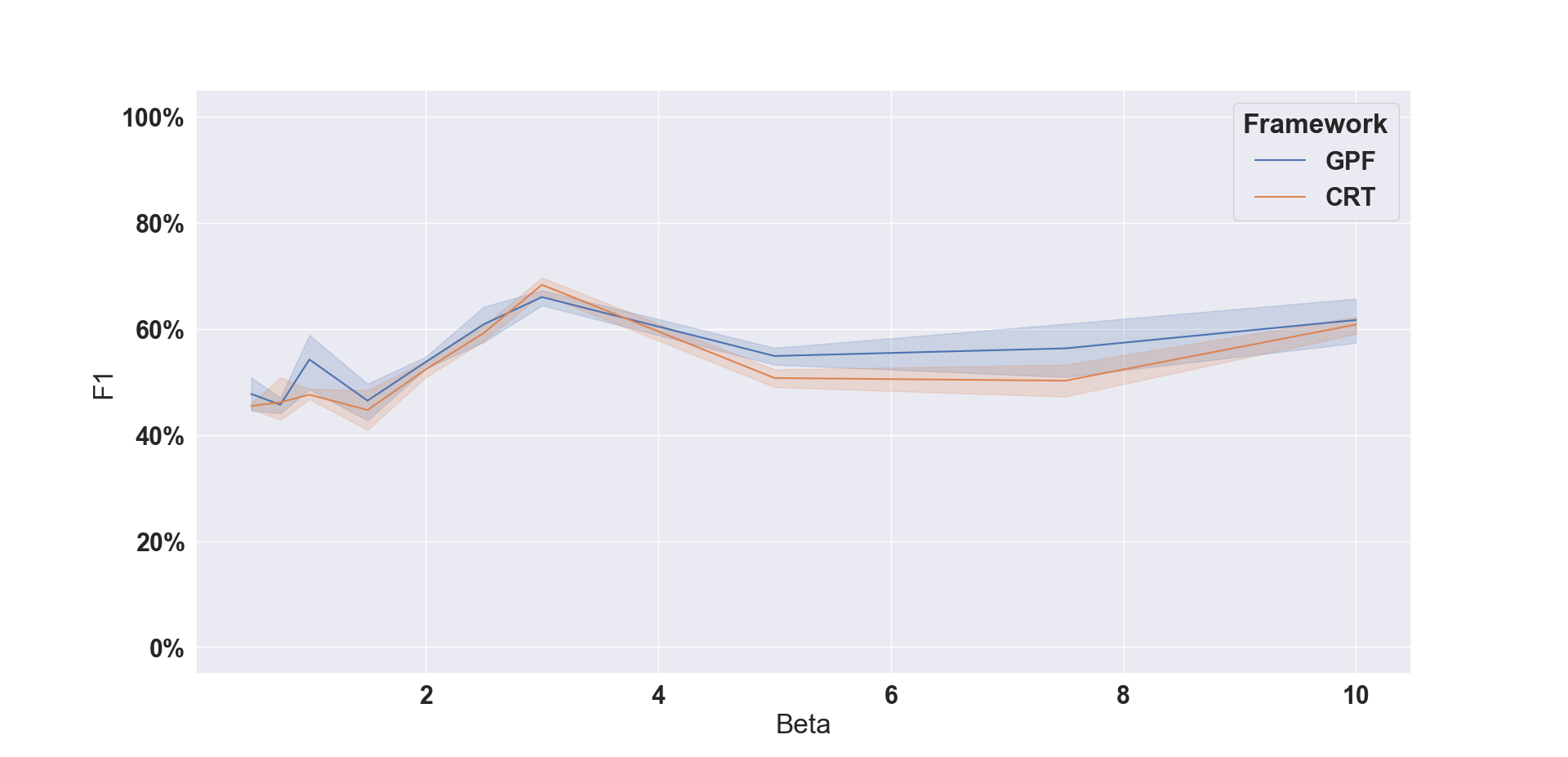}
\caption{$F_{1}$ vs $\beta$}\label{fig.c}
\end{subfigure}%
\begin{subfigure}{.50\linewidth}
\includegraphics[width=1.15\linewidth]{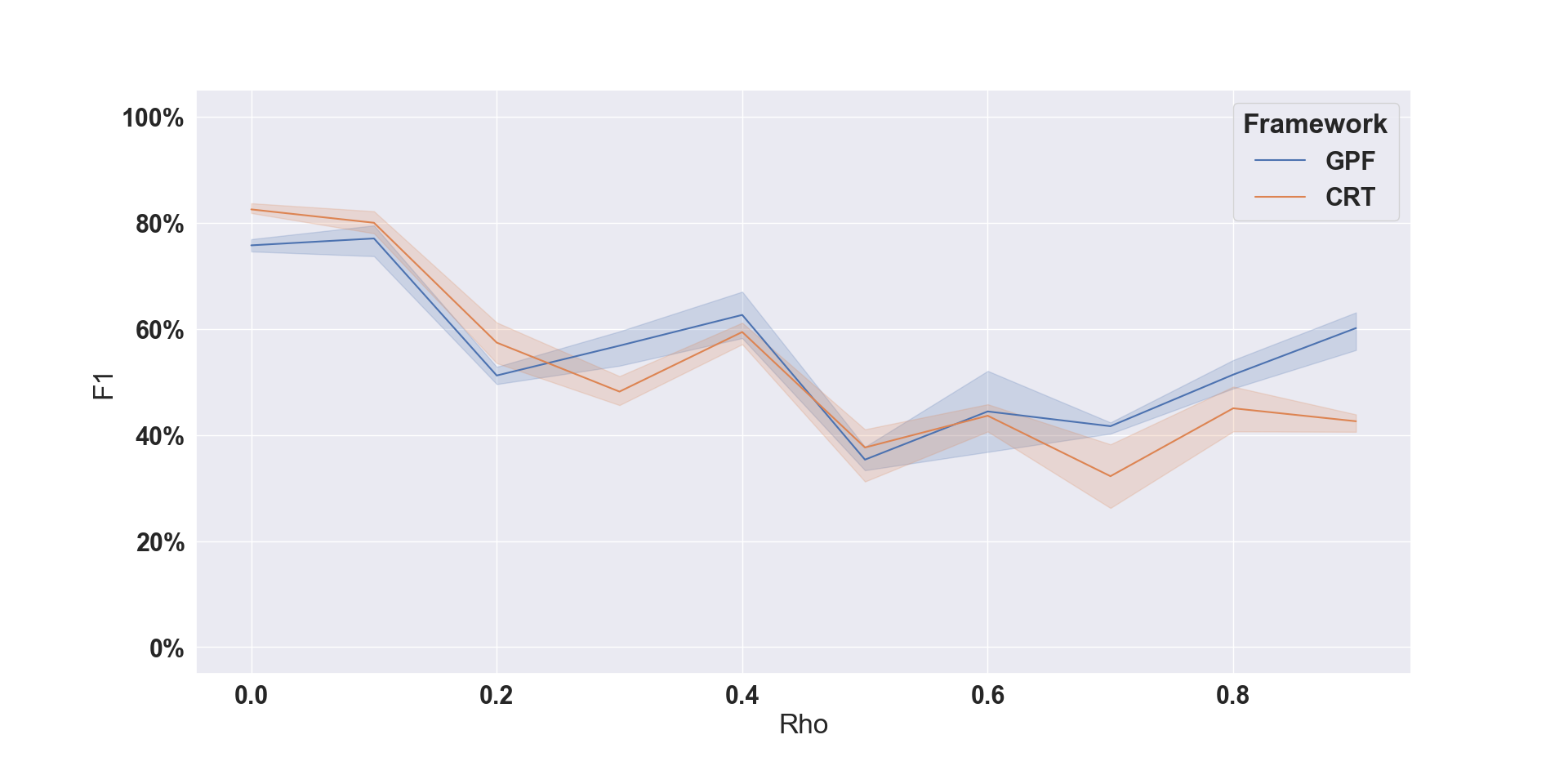}
\caption{$F_{1}$ vs $\rho$}\label{fig.f}
\end{subfigure}%
\\
\begin{subfigure}{.50\linewidth}
\includegraphics[width=1.15\linewidth]{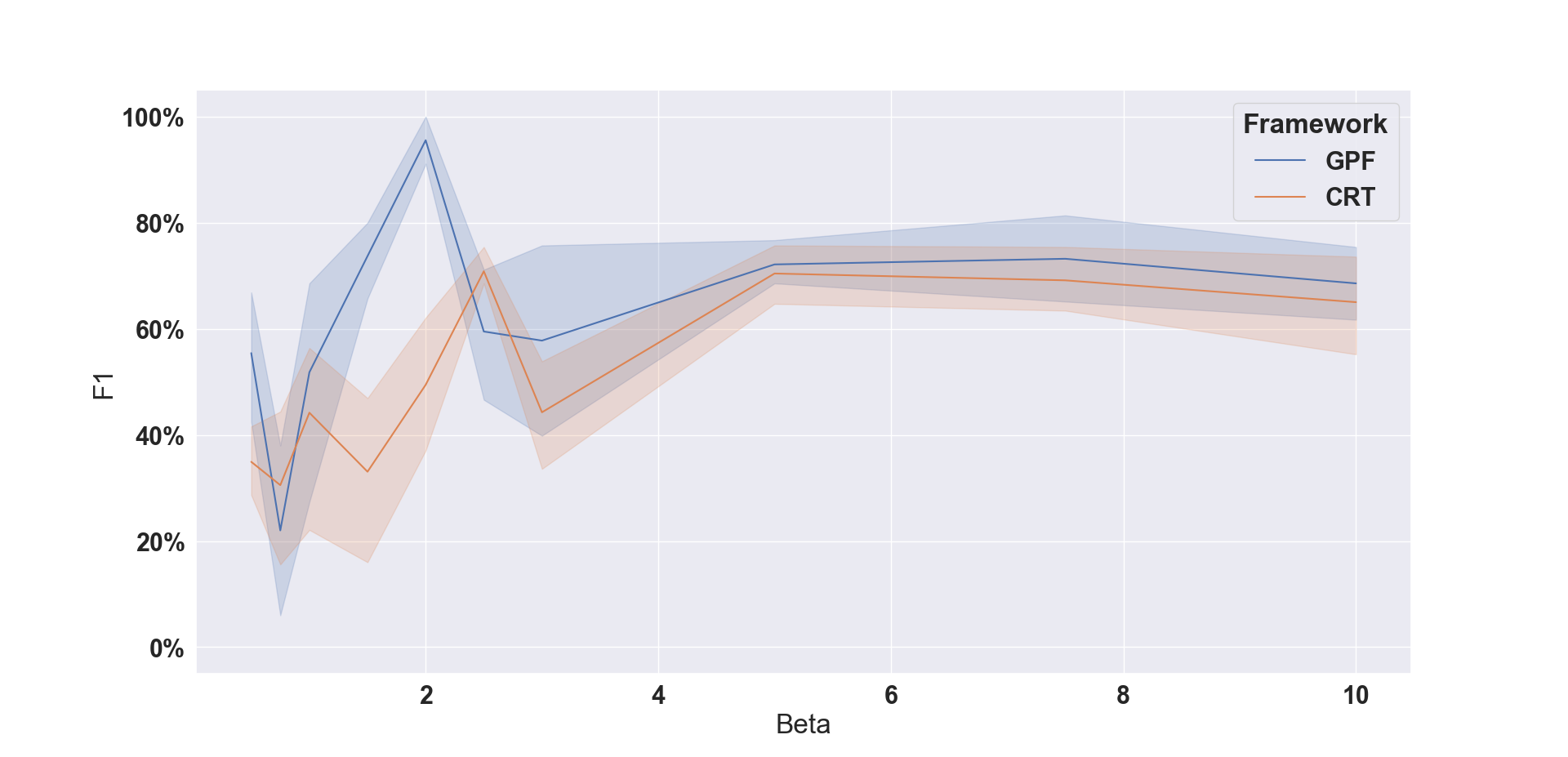}
\caption{$F_{1}$ vs $\beta$}\label{fig.c}
\end{subfigure}%
\begin{subfigure}{.50\linewidth}
\includegraphics[width=1.15\linewidth]{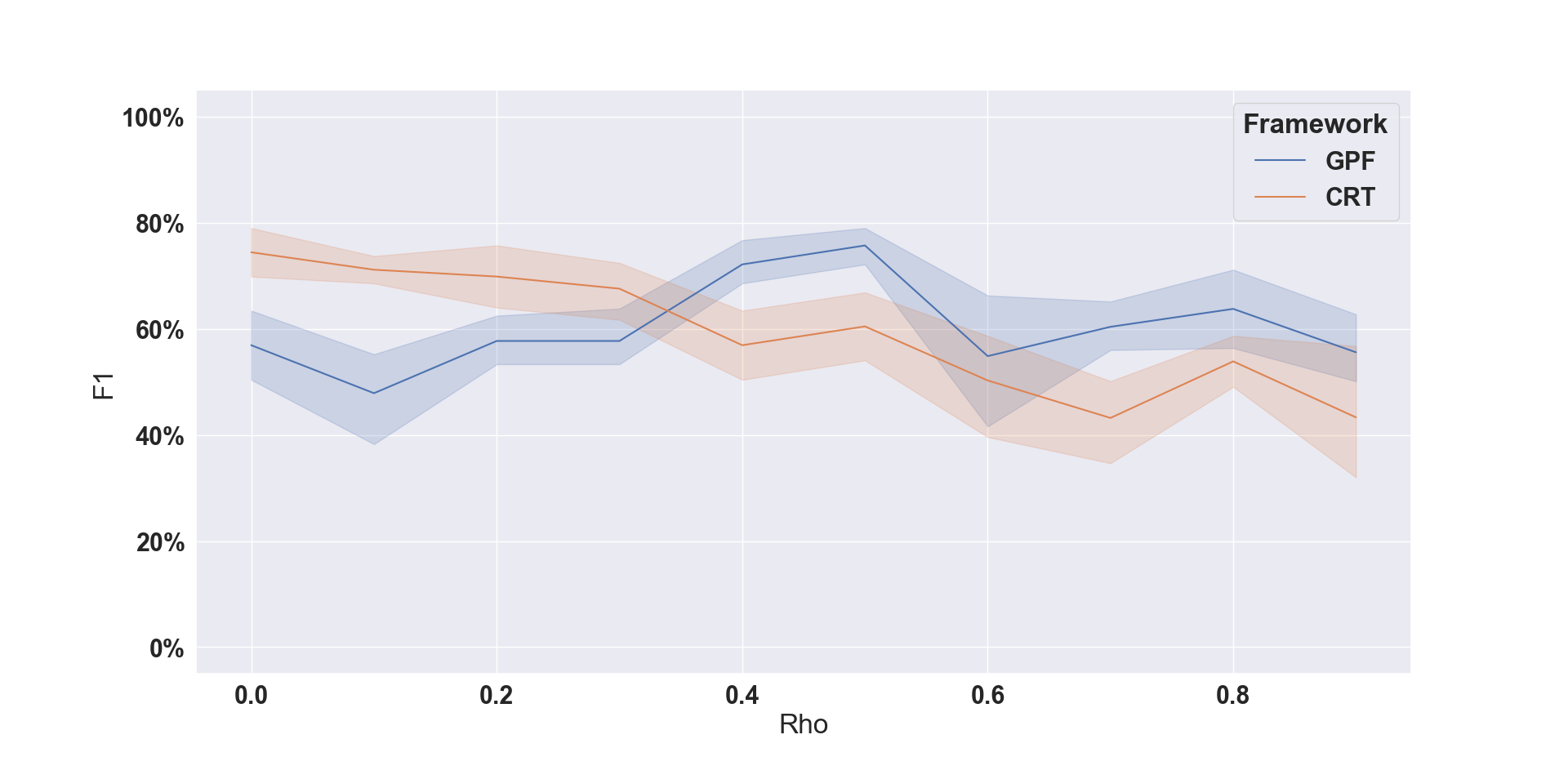}
\caption{$F_{1}$ vs $\rho$}\label{fig.f}
\end{subfigure}%
\caption{$F_{1}$ for CRT and GPF against signal strengths $\beta$ in the left column, and against variable correlation $\rho$ in the right with logistic data in the top row and step-wise data on the bottom.}\label{fig:logisticComp}
\vskip -0.15in
\end{figure}

For the binomial transformed data,
a modified LCD, as suggested in \citep{candes2018panning}, was used as the statistic.
Subset-GPF slightly outperformed CRT in terms of $F_{1}$ for all $\beta$ and most $\rho$ in Figure~\ref{fig:logisticComp} a and b.
Finally, for the step-wise data, the tree feature importance difference was used as the test statistic.
The $F_{1}$ for Subset-GPF is slightly higher than CRT for various $\beta$ and for the more challenging high correlation settings of $\rho$. 
We attempted to run HRT for the binomial and step-wise data, but was unsuccessful in determining significant different variables using the test accuracy as the HRT model score.

\subsection{Complexity and run times}
The training model complexities are shown in Table~\ref{table:trainingComplexity}.
\begin{table}[t]
\begin{minipage}{.495\textwidth}
\begin{center}
\caption{Training Complexity}
\label{table:trainingComplexity}
\end{center}
\begin{small}
\begin{tabular}{lcr}
\toprule
Model & $O(\cdot)$  \\
\midrule
Lasso & $O(D^{3} + N \times D^{2})$ \\
Logistic & $O(D \times C \times N \times E )$  \\
XGBoost & $O(T \times H \times D \times N \times \log{N})$ \\
\bottomrule
\end{tabular}
\end{small}
\end{minipage}
\begin{minipage}{.495\textwidth}
\begin{center}
\caption{Run Times in seconds}
\label{table:trainingRunTimes}
\end{center}
\begin{small}
\begin{tabular}{lcr}
\toprule
Model & CRT & Subset-GPF  \\
\midrule
Lasso & $1955 \pm 81$ & $229 \pm 9$ \\
Logistic & $10320 \pm 448$ & $772 \pm 37$ \\
XGBoost & $3277 \pm 134 $ & $ 1300 \pm 98 $ \\
\bottomrule
\end{tabular}
\end{small}
\end{minipage}
\vskip -0.1in
\end{table}
Training the Lasso model is cubic in the number of variables $D$ \citep{efron2004least}.
The logistic regression model, with no intercept $C$ target classes, and maximum stopping epoch $E$, can be trained in $O(DCNE)$.
For our experiments with logistic regression with L1 penalty, $E=2000$.
Finally, XGBoost takes $O(TH DN\log{N})$ \citep{chen2016xgboost}, where $T$ is the number and $H$ the height of the trees.
For our experiments using XGBoost, $T=100$ and $H=3$.
Since $D=P=400$ for CRT, while $D=K=\left \lfloor{\sqrt{N} } \right \rfloor= 15$ for Subset-GPF in our synthetic data experiments, Subset-GPF is faster than CRT.

We measured running times for CRT and Subset-GPF for one variable $X_{1}$ clocked on the same CPU over 10 replications, and present them in Table~\ref{table:trainingRunTimes}.
These training times include the conditional sampling for CRT and the random permutations for Subset-GPF.

As expected, Subset-GPF was substantially quicker to train for all three methods as it had lower number of variables $D=K=15$, whereas CRT had to train over $D=P=400$.
We do note that training run times were not as different as the ratio of training complexity would suggest.
This is partly due to an efficient conditional sampling based on the conditional multivariate Gaussian distribution implemented for CRT, 
while Subset-GPF used random shuffle permutations.
These run times are for CRT or Subset-GPF on one variable,
and would be expensive if all variables were run sequentially.
Sequential run times can be reduced by parallelizing Subset-GPF and CRT in the cloud.

\subsection{Allen Brain data results}
\begin{figure}[!h]
  \begin{minipage}[b]{.48\linewidth}
    \centering
    \includegraphics[scale=.32]{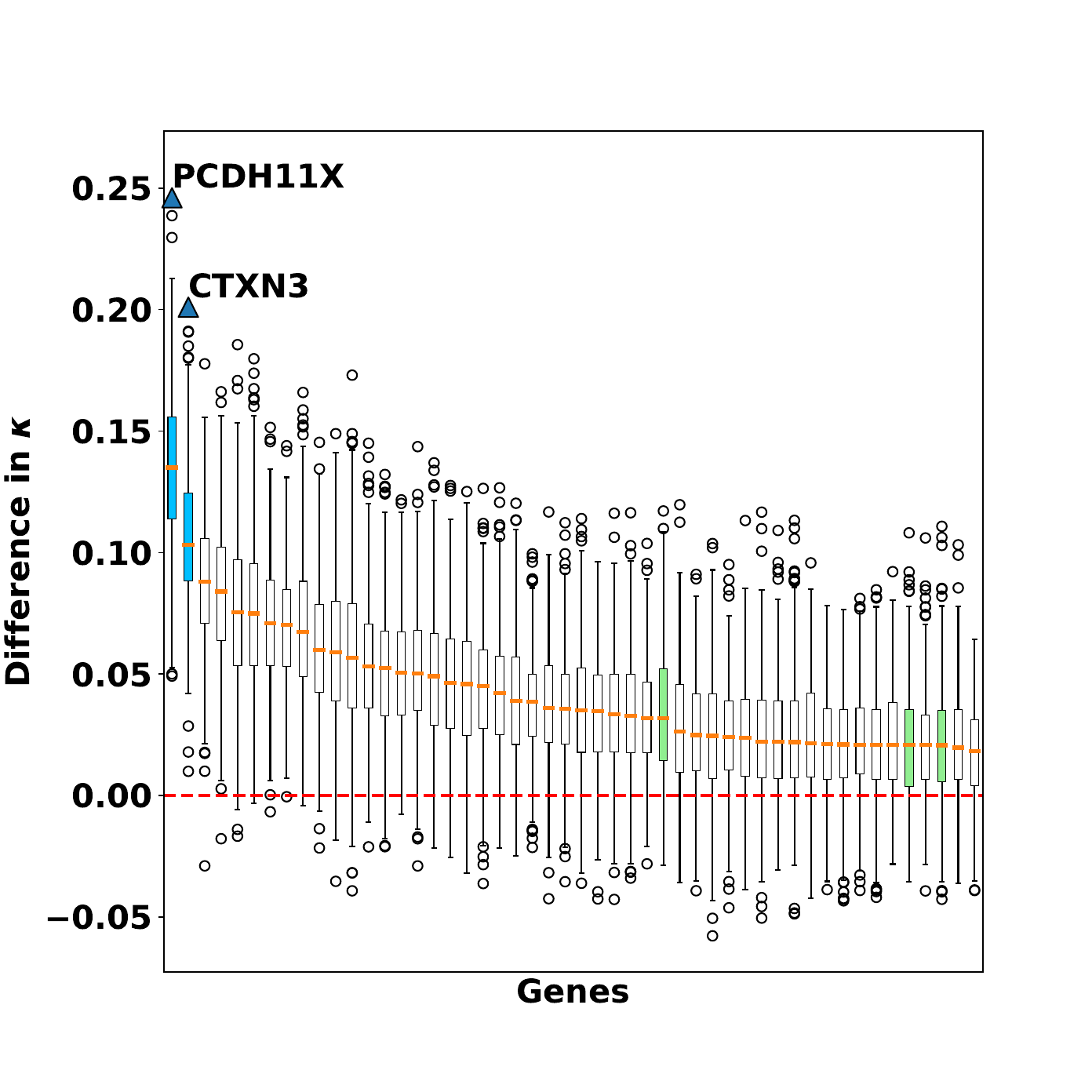}
    \vskip -0.25in
    \caption{First $50$ genes ordered by median $\Delta(\kappa_{\Phi_{x},\cdot})$. Significant genes at $\alpha=\frac{0.05}{50000}$ are shaded blue. Green shaded boxes indicate randomly selected genes.}
    \label{fig:allen_paired_kappa_boxplot}
  \end{minipage}
  \hfill
  \begin{minipage}[b]{.48\linewidth}
    \centering
    \includegraphics[scale=.32]{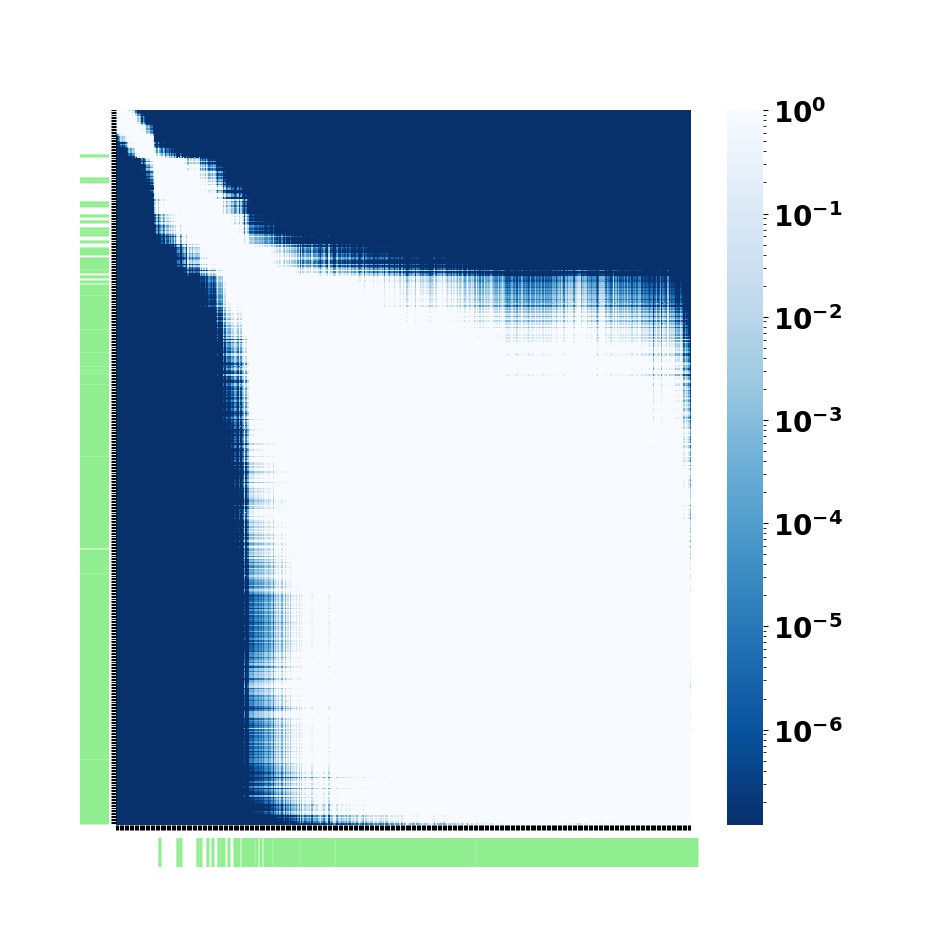}
    \vskip -0.2in
    \caption{Pairwise Conover p-values of the ordered $\Delta(\kappa_{\Phi_{x}, \cdot})$ for all $500$ genes. Green bars indicate randomly selected genes. Genes ordered in the same fashion as Figure~\ref{fig:allen_paired_kappa_boxplot} and \ref{fig:allen_paired_kappa_boxplot_all}}
    \label{fig:allen_paired_kappa_conover}
  \end{minipage}
\end{figure}
We ran Subset-GPF (Algorithm~\ref{alg:subsetGPF})
with $f$ chosen to be XGBoost and $M=500$ genes of interest,
since we cannot run CRT or HRT as we do not know the marginal distribution of the genes. 
The genes of interest were chosen to be the top $100$ genes in terms of feature importance from XGBoost, run on the entire variable matrix $\mathbf{X}$,
and $400$ randomly selected genes were chosen as well.
Including the top $100$ genes from XGBoost allowed us to analyze genes that would be expected to be in the tail distribution in terms of model contribution.
The number of runs for each gene was set to be $R=1000$.
For each $R$, a subsample of $K-1=49$ genes were used in combination with the tested gene $\mathbf{x_{j_{m}}}$.
In each run, the data was split randomly 80\%-20\% in terms of training and test.
XGBoost was trained on the training portion and evaluated on the test.
$\kappa$ agreement were computed for test predictions as this is a multi-class classification problem.
Furthermore, the delta in $\kappa$ on the test dataset for Subset-GPF was used as the test statistic.

Figure~\ref{fig:allen_paired_kappa_boxplot} shows the ordered differenced $\kappa$ null distributions $\Delta(\kappa_{\Phi_{x},\cdot})$, for $50$ genes from Subset-GPF;
the entire $500$ gene plot is given in Supplemental Figure~\ref{fig:allen_paired_kappa_boxplot_all}.
Two genes, \emph{PCDH11X} and \emph{CTXN3}, showed statistically significant effects after Bonferroni correction.
We note that a few randomly selected genes, shaded green, were among the top $50$ genes by $\Delta(\kappa_{\Phi_{x},\cdot})$, suggesting that tail distribution of the top 100 genes is similar to random genes.

The Conover test with Bonferroni adjustment \citep{conover1979multiple} was used to test differences in the differenced null distributions $\Delta(\kappa_{\Phi_{x}, \cdot})$ between every pairwise gene combination for all $500$ genes (Figure~\ref{fig:allen_paired_kappa_conover}). 
A small cluster of $14$ genes (upper left) were found to be statistically indistinguishable from \emph{PCDH11X} and \emph{CTXN3} 
and were a XGBoost model was trained only on these genes to evaluate their sufficiency. 
For reference, XGBoost was run on these $14$ genes without permutation, and achieved $\kappa=0.85$, weighted $F1=0.89$ and $recall=0.89$.
These metrics shows that the $14$ genes were sufficient for classification.

The three genes, \emph{PCDH11x}, \emph{CTXN3}, and \emph{BHLHE22}, have been shown to be differentially expressed in targeted assays in the medical literature, but to our knowledge, not before in whole transcriptomic data.  
\emph{PCDH11X} has been shown both by real-time reverse transcription-polymerase chain reaction amplification assays \citep{Blanco2000-hw, Ahn2010-vt} 
and by immunohistochemistry \citep{Priddle2013-ky} to be differentially expressed in adult human brains. 
\citet{Wang2007-uz} showed that \emph{CTXN3} is highly conserved among vertebrates with brain-specific expression. 
In \emph{BHLB5 -/-} mice, \citep{Ross2012-ep} showed agency of the hippocampal commissure, anterior commissure, and corpus callosum. 
In addition, \emph{BHLB5} has been shown to be a post-mitotic regulator of the neocortex including the frontal lobe. \citep{Joshi2008-xx, Bedogni2010-lq} 
Interestingly, variants in both \emph{PCDH11X} and \emph{CTXN3} have been associated with schizophrenia \citep{Levchenko2014-lo,Panichareon2012-ao, Potkin2009-lj, Sery2015-uh}. As a baseline comparison of another permutation framework, multiclass SAMseq \citep{li2013finding} was run on the same dataset using $1000$ permutations. 
No genes were found to be statistically significant with the default false discovery rate of 20\% (Figure~\ref{fig:allen_samseq}). 

\section{Conclusion}
We presented a generalized permutation framework, GPF, that is able to test for the significance of variables and their interactions in supervised models. 
The framework was applied to synthetic datasets and a high dimensional, real-world RNA expression dataset.
Using GPF, we were able to identify significant variable interactions in a model-independent fashion for the synthetic datasets. 
In addition, we found genes in whole transcriptome data to be differentially expressed by brain regions. 
These genes have been previously confirmed in targeted biochemical and immunohistochemical assays to be differentially enriched across brain regions.
Our experiments show promising results for the applicability of GPF in high-dimensional datasets with supervised models and its ability to elucidate statistical relationships among the variables. 

\bibliography{perm}

\begin{thebibliography}{40}
\providecommand{\natexlab}[1]{#1}
\providecommand{\url}[1]{\texttt{#1}}
\expandafter\ifx\csname urlstyle\endcsname\relax
  \providecommand{\doi}[1]{doi: #1}\else
  \providecommand{\doi}{doi: \begingroup \urlstyle{rm}\Url}\fi

\bibitem[Ahn et~al.(2010)Ahn, Huh, Kim, Ha, Kim, Lee, and Kim]{Ahn2010-vt}
Kung Ahn, Jae-Won Huh, Dae-Soo Kim, Hong-Seok Ha, Yun-Ji Kim, Ja-Rang Lee, and
  Heui-Soo Kim.
\newblock Quantitative analysis of alternative transcripts of human {PCDH11X/Y}
  genes.
\newblock \emph{Am. J. Med. Genet. B Neuropsychiatr. Genet.}, 153B\penalty0
  (3):\penalty0 736--744, April 2010.

\bibitem[Altmann et~al.(2010)Altmann, Tolo{\c{s}}i, Sander, and
  Lengauer]{altmann2010permutation}
Andr{\'e} Altmann, Laura Tolo{\c{s}}i, Oliver Sander, and Thomas Lengauer.
\newblock Permutation importance: a corrected feature importance measure.
\newblock \emph{Bioinformatics}, 26\penalty0 (10):\penalty0 1340--1347, 2010.

\bibitem[Bedogni et~al.(2010)Bedogni, Hodge, Elsen, Nelson, Daza, Beyer,
  Bammler, Rubenstein, and Hevner]{Bedogni2010-lq}
Francesco Bedogni, Rebecca~D Hodge, Gina~E Elsen, Branden~R Nelson, Ray A~M
  Daza, Richard~P Beyer, Theo~K Bammler, John L~R Rubenstein, and Robert~F
  Hevner.
\newblock Tbr1 regulates regional and laminar identity of postmitotic neurons
  in developing neocortex.
\newblock \emph{Proc. Natl. Acad. Sci. U. S. A.}, 107\penalty0 (29):\penalty0
  13129--13134, July 2010.

\bibitem[Berrett et~al.(2020)Berrett, Wang, Barber, and
  Samworth]{berrett2020conditional}
Thomas~B Berrett, Yi~Wang, Rina~Foygel Barber, and Richard~J Samworth.
\newblock The conditional permutation test for independence while controlling
  for confounders.
\newblock \emph{Journal of the Royal Statistical Society: Series B (Statistical
  Methodology)}, 82\penalty0 (1):\penalty0 175--197, 2020.

\bibitem[Blanco et~al.(2000)Blanco, Sargent, Boucher, Mitchell, and
  Affara]{Blanco2000-hw}
P~Blanco, C~A Sargent, C~A Boucher, M~Mitchell, and N~A Affara.
\newblock Conservation of {PCDHX} in mammals; expression of human {X/Y} genes
  predominantly in brain.
\newblock \emph{Mamm. Genome}, 11\penalty0 (10):\penalty0 906--914, October
  2000.

\bibitem[Breiman et~al.(2017)Breiman, Friedman, Olshen, and
  Stone]{breiman2017classification}
Leo Breiman, Jerome~H Friedman, Richard~A Olshen, and Charles~J Stone.
\newblock \emph{Classification and regression trees}.
\newblock Routledge, 2017.

\bibitem[Candes et~al.(2018)Candes, Fan, Janson, and Lv]{candes2018panning}
Emmanuel Candes, Yingying Fan, Lucas Janson, and Jinchi Lv.
\newblock Panning for gold:‘model-x’knockoffs for high dimensional
  controlled variable selection.
\newblock \emph{Journal of the Royal Statistical Society: Series B (Statistical
  Methodology)}, 80\penalty0 (3):\penalty0 551--577, 2018.

\bibitem[Chen and Guestrin(2016)]{chen2016xgboost}
Tianqi Chen and Carlos Guestrin.
\newblock Xgboost: A scalable tree boosting system.
\newblock In \emph{Proceedings of the 22nd acm sigkdd international conference
  on knowledge discovery and data mining}, pages 785--794, 2016.

\bibitem[Conover and Iman(1979)]{conover1979multiple}
William~Jay Conover and Ronald~L Iman.
\newblock On multiple-comparisons procedures.
\newblock \emph{Los Alamos Sci. Lab. Tech. Rep. LA-7677-MS}, pages 1--14, 1979.

\bibitem[Efron et~al.(2004)Efron, Hastie, Johnstone, and
  Tibshirani]{efron2004least}
Bradley Efron, Trevor Hastie, Iain Johnstone, and Robert Tibshirani.
\newblock Least angle regression.
\newblock 2004.

\bibitem[Fisher(1949)]{fisher1949design}
Ronald~A Fisher.
\newblock The design of experiments.
\newblock 1949.

\bibitem[Golland and Fischl(2003)]{golland2003permutation}
Polina Golland and Bruce Fischl.
\newblock Permutation tests for classification: towards statistical
  significance in image-based studies.
\newblock In \emph{Biennial international conference on information processing
  in medical imaging}, pages 330--341. Springer, 2003.

\bibitem[Good(2013)]{good2013permutation}
Phillip Good.
\newblock \emph{Permutation tests: a practical guide to resampling methods for
  testing hypotheses}.
\newblock Springer Science \& Business Media, 2013.

\bibitem[Goodfellow et~al.(2016)Goodfellow, Bengio, and
  Courville]{goodfellow2016deep}
Ian Goodfellow, Yoshua Bengio, and Aaron Courville.
\newblock \emph{Deep learning}.
\newblock MIT press, 2016.

\bibitem[Hoeffding(1952)]{hoeffding1952large}
Wassily Hoeffding.
\newblock The large-sample power of tests based on permutations of
  observations.
\newblock \emph{The Annals of Mathematical Statistics}, pages 169--192, 1952.

\bibitem[Joshi et~al.(2008)Joshi, Molyneaux, Feng, Xie, Macklis, and
  Gan]{Joshi2008-xx}
Pushkar~S Joshi, Bradley~J Molyneaux, Liang Feng, Xiaoling Xie, Jeffrey~D
  Macklis, and Lin Gan.
\newblock Bhlhb5 regulates the postmitotic acquisition of area identities in
  layers {II-V} of the developing neocortex.
\newblock \emph{Neuron}, 60\penalty0 (2):\penalty0 258--272, October 2008.

\bibitem[Kumar et~al.(2020)Kumar, Venkatasubramanian, Scheidegger, and
  Friedler]{kumar2020problems}
I~Elizabeth Kumar, Suresh Venkatasubramanian, Carlos Scheidegger, and Sorelle
  Friedler.
\newblock Problems with shapley-value-based explanations as feature importance
  measures.
\newblock In \emph{International Conference on Machine Learning}, pages
  5491--5500. PMLR, 2020.

\bibitem[Lehr(1992)]{lehr1992sixteen}
Robert Lehr.
\newblock Sixteen s-squared over d-squared: A relation for crude sample size
  estimates.
\newblock \emph{Statistics in medicine}, 11\penalty0 (8):\penalty0 1099--1102,
  1992.

\bibitem[Levchenko et~al.(2014)Levchenko, Davtian, Petrova, and
  Malashichev]{Levchenko2014-lo}
Anastasia Levchenko, Stepan Davtian, Natalia Petrova, and Yegor Malashichev.
\newblock Sequencing of five left-right cerebral asymmetry genes in a cohort of
  schizophrenia and schizotypal disorder patients from russia.
\newblock \emph{Psychiatr. Genet.}, 24\penalty0 (2):\penalty0 75--80, April
  2014.

\bibitem[Li and Dewey(2011)]{li2011rsem}
Bo~Li and Colin~N Dewey.
\newblock Rsem: accurate transcript quantification from rna-seq data with or
  without a reference genome.
\newblock \emph{BMC bioinformatics}, 12\penalty0 (1):\penalty0 323, 2011.

\bibitem[Li and Tibshirani(2013)]{li2013finding}
Jun Li and Robert Tibshirani.
\newblock Finding consistent patterns: a nonparametric approach for identifying
  differential expression in rna-seq data.
\newblock \emph{Statistical methods in medical research}, 22\penalty0
  (5):\penalty0 519--536, 2013.

\bibitem[Lindgren et~al.(1996)Lindgren, Hansen, Karcher, Sj{\"o}str{\"o}m, and
  Eriksson]{lindgren1996model}
Fredrik Lindgren, Bj{\"o}rn Hansen, Walter Karcher, Michael Sj{\"o}str{\"o}m,
  and Lennart Eriksson.
\newblock Model validation by permutation tests: applications to variable
  selection.
\newblock \emph{Journal of Chemometrics}, 10\penalty0 (5-6):\penalty0 521--532,
  1996.

\bibitem[Love et~al.(2014)Love, Huber, and Anders]{love2014moderated}
Michael~I Love, Wolfgang Huber, and Simon Anders.
\newblock Moderated estimation of fold change and dispersion for rna-seq data
  with deseq2.
\newblock \emph{Genome biology}, 15\penalty0 (12):\penalty0 550, 2014.

\bibitem[Lundberg and Lee(2017)]{lundberg2017unified}
Scott~M Lundberg and Su-In Lee.
\newblock A unified approach to interpreting model predictions.
\newblock In \emph{Advances in neural information processing systems}, pages
  4765--4774, 2017.

\bibitem[Miller et~al.(2017)Miller, Guillozet-Bongaarts, Gibbons, Postupna,
  Renz, Beller, Sunkin, Ng, Rose, Smith, et~al.]{miller2017neuropathological}
Jeremy~A Miller, Angela Guillozet-Bongaarts, Laura~E Gibbons, Nadia Postupna,
  Anne Renz, Allison~E Beller, Susan~M Sunkin, Lydia Ng, Shannon~E Rose,
  Kimberly~A Smith, et~al.
\newblock Neuropathological and transcriptomic characteristics of the aged
  brain.
\newblock \emph{Elife}, 6:\penalty0 e31126, 2017.

\bibitem[Panichareon et~al.(2012)Panichareon, Nakayama, Iwamoto,
  Thurakitwannakarn, and Sukhumsirichart]{Panichareon2012-ao}
Benjaporn Panichareon, Kazuhiro Nakayama, Sadahiko Iwamoto, Wanpen
  Thurakitwannakarn, and Wasana Sukhumsirichart.
\newblock Association of {CTXN3-SLC12A2} polymorphisms and schizophrenia in a
  thai population.
\newblock \emph{Behav. Brain Funct.}, 8:\penalty0 27, May 2012.

\bibitem[Potkin et~al.(2009)Potkin, Turner, Guffanti, Lakatos, Fallon, Nguyen,
  Mathalon, Ford, Lauriello, Macciardi, and {FBIRN}]{Potkin2009-lj}
Steven~G Potkin, Jessica~A Turner, Guia Guffanti, Anita Lakatos, James~H
  Fallon, Dana~D Nguyen, Daniel Mathalon, Judith Ford, John Lauriello, Fabio
  Macciardi, and {FBIRN}.
\newblock A genome-wide association study of schizophrenia using brain
  activation as a quantitative phenotype.
\newblock \emph{Schizophr. Bull.}, 35\penalty0 (1):\penalty0 96--108, January
  2009.

\bibitem[Priddle and Crow(2013)]{Priddle2013-ky}
Thomas~H Priddle and Tim~J Crow.
\newblock Protocadherin {11X/Y} a human-specific gene pair: an
  immunohistochemical survey of fetal and adult brains.
\newblock \emph{Cereb. Cortex}, 23\penalty0 (8):\penalty0 1933--1941, August
  2013.

\bibitem[Ribeiro et~al.(2016)Ribeiro, Singh, and Guestrin]{ribeiro2016should}
Marco~Tulio Ribeiro, Sameer Singh, and Carlos Guestrin.
\newblock " why should i trust you?" explaining the predictions of any
  classifier.
\newblock In \emph{Proceedings of the 22nd ACM SIGKDD international conference
  on knowledge discovery and data mining}, pages 1135--1144, 2016.

\bibitem[Ross et~al.(2012)Ross, McCord, Jung, Atan, Mok, Hemberg, Kim,
  Salogiannis, Hu, Cohen, Lin, Harrar, McInnes, and Greenberg]{Ross2012-ep}
Sarah~E Ross, Alejandra~E McCord, Cynthia Jung, Denize Atan, Stephanie~I Mok,
  Martin Hemberg, Tae-Kyung Kim, John Salogiannis, Linda Hu, Sonia Cohen,
  Yingxi Lin, Dana Harrar, Roderick~R McInnes, and Michael~E Greenberg.
\newblock Bhlhb5 and prdm8 form a repressor complex involved in neuronal
  circuit assembly.
\newblock \emph{Neuron}, 73\penalty0 (2):\penalty0 292--303, January 2012.

\bibitem[Scheffe(1943)]{scheffe1943statistical}
Henry Scheffe.
\newblock Statistical inference in the non-parametric case.
\newblock \emph{The Annals of Mathematical Statistics}, 14\penalty0
  (4):\penalty0 305--332, 1943.

\bibitem[Selvaraju et~al.(2017)Selvaraju, Cogswell, Das, Vedantam, Parikh, and
  Batra]{selvaraju2017grad}
Ramprasaath~R Selvaraju, Michael Cogswell, Abhishek Das, Ramakrishna Vedantam,
  Devi Parikh, and Dhruv Batra.
\newblock Grad-cam: Visual explanations from deep networks via gradient-based
  localization.
\newblock In \emph{Proceedings of the IEEE international conference on computer
  vision}, pages 618--626, 2017.

\bibitem[{\v S}er{\'y} et~al.(2015){\v S}er{\'y}, Lochman, Povov{\'a}, Janout,
  Plesn{\'\i}k, and Balcar]{Sery2015-uh}
Omar {\v S}er{\'y}, Jan Lochman, Jana Povov{\'a}, Vladim{\'\i}r Janout, Ji{\v
  r}{\'\i} Plesn{\'\i}k, and Vladimir~J Balcar.
\newblock Association between 5q23.2-located polymorphism of {CTXN3} gene
  (cortexin 3) and schizophrenia in {European-Caucasian} males; implications
  for the aetiology of schizophrenia.
\newblock \emph{Behav. Brain Funct.}, 11:\penalty0 10, March 2015.

\bibitem[Simonyan et~al.(2013)Simonyan, Vedaldi, and
  Zisserman]{simonyan2013deep}
Karen Simonyan, Andrea Vedaldi, and Andrew Zisserman.
\newblock Deep inside convolutional networks: Visualising image classification
  models and saliency maps.
\newblock \emph{arXiv preprint arXiv:1312.6034}, 2013.

\bibitem[Strobl et~al.(2008)Strobl, Boulesteix, Kneib, Augustin, and
  Zeileis]{strobl2008conditional}
Carolin Strobl, Anne-Laure Boulesteix, Thomas Kneib, Thomas Augustin, and Achim
  Zeileis.
\newblock Conditional variable importance for random forests.
\newblock \emph{BMC bioinformatics}, 9\penalty0 (1):\penalty0 307, 2008.

\bibitem[Tansey et~al.(2022)Tansey, Veitch, Zhang, Rabadan, and
  Blei]{tansey2022holdout}
Wesley Tansey, Victor Veitch, Haoran Zhang, Raul Rabadan, and David~M Blei.
\newblock The holdout randomization test for feature selection in black box
  models.
\newblock \emph{Journal of Computational and Graphical Statistics}, 31\penalty0
  (1):\penalty0 151--162, 2022.

\bibitem[Van~Belle(2011)]{van2011statistical}
Gerald Van~Belle.
\newblock \emph{Statistical rules of thumb}, volume 699.
\newblock John Wiley \& Sons, 2011.

\bibitem[Wald and Wolfowitz(1944)]{wald1944statistical}
Abraham Wald and Jacob Wolfowitz.
\newblock Statistical tests based on permutations of the observations.
\newblock \emph{The Annals of Mathematical Statistics}, 15\penalty0
  (4):\penalty0 358--372, 1944.

\bibitem[Wang et~al.(2007)Wang, Chang, Guo, and Li]{Wang2007-uz}
Hai~Tao Wang, Ji~Wu Chang, Zhi Guo, and Bao~Guo Li.
\newblock In silico-initiated cloning and molecular characterization of
  cortexin 3, a novel human gene specifically expressed in the kidney and
  brain, and well conserved in vertebrates.
\newblock \emph{Int. J. Mol. Med.}, 20\penalty0 (4):\penalty0 501--510, October
  2007.

\bibitem[Williamson et~al.(2017)Williamson, Gilbert, Simon, and
  Carone]{williamson2017nonparametric}
Brian~D Williamson, Peter~B Gilbert, Noah Simon, and Marco Carone.
\newblock Nonparametric variable importance assessment using machine learning
  techniques.
\newblock 2017.

\end{thebibliography}
\bibliographystyle{plainnat}

\newpage
\appendix
\onecolumn
\renewcommand\thefigure{\thesection.\arabic{figure}} 
\setcounter{figure}{0}  
\renewcommand{\thealgorithm}{A\arabic{algorithm}}

\section{Conditional Randomization Test Algorithm }
\setcounter{algorithm}{0}
\begin{algorithm}[ht]
   \caption{Conditional Randomization Test}
   \label{alg:CRT_algo}
\belowdisplayskip=-5pt 
\begin{algorithmic}[1]
   \STATE {\bfseries Input:} response $\mathbf{y}=\{y_{i}\}_{i=1}^{N}$, variables $\mathbf{X}= \{ x_{ij} \}_{i=1,j=1}^{N,P}$, the marginal distribution $\mathcal{F}(\mathbf{X})$, a supervised model $f$, and feature importance statistic $T_{j}(\mathbf{X}, \mathbf{y})$
   \FOR{$r=1$ to $R$ do}
   	\FOR{$j=1$ to $P$ do}
	   	\STATE Sample 
		   	\begin{equation}
		   		X_{ij}^{(r)} \sim \mathcal{F}(X_{ij} | \{X_{ip} \}_{p \neq j} ), \; \forall i=1,...,N \label{eq:simConditional}
			\end{equation}
	   	\STATE Replace column $j$ in $\mathbf{X}$ with $X_{j}^{(r)}$ to get $\mathbf{X}^{(r)}$
	   	\STATE Compute $\mathbbm{1}(T_{j}(\mathbf{X}^{(r)}, \mathbf{y}) \leq T_{j}(\mathbf{X}, \mathbf{y}))$
   	\ENDFOR
   \ENDFOR
   \STATE {\bfseries Output}: One-sided p-values for $j=1,...,P$: 
   	\begin{equation}
   		p_{j} = \frac{1}{R+1}\left[ 1 + \sum_{r=1}^{R} \mathbbm{1}(T_{j}(\mathbf{X}^{(r)}, \mathbf{y}) \leq T_{j}(\mathbf{X}, \mathbf{y})) \right] \label{eq:crt_pvalue}
	\end{equation}
\end{algorithmic}
\end{algorithm}
In \citet{candes2018panning}, the authors propose using coefficient of variable $X_{j}$ from a fitted Lasso model $f$, with a penalty $\lambda$ for linear-Gaussian datasets as the test statistic, so that $T_{j}(\mathbf{X}, \mathbf{y})=\hat{b}_{j}(\lambda)$.
Then the one sided p-value test in Equation~\ref{eq:crt_pvalue} simplifies to:
\begin{align}
	W_{j}^{(r)} &= |\hat{b}_{j}(\lambda)| - |\hat{b}_{j}^{(r)}(\lambda)| \\ \label{eq:lcd}
	p_{j} &= 1- \frac{1}{R+1}\left[ 1 + \sum_{r=1}^{R} \mathbbm{1}(W_{j}^{(r)} > 0) \right]
\end{align}
Here $\hat{b}_{j}^{(r)}(\lambda)$ denotes the coefficient of the perturbed variable $x_{j}^{(r)}$ in a fitted Lasso model with a penalty $\lambda$.
\citet{candes2018panning} refers to Equation~\ref{eq:lcd} as the Lasso Coefficient Difference (LCD).
The intuition is that if variable $X_{j}$ contributes to predicting $\mathbf{y}$ under a linear model $f$, then its coefficient, $\hat{b}_{j}(\lambda)$, will be nonzero and larger than the coefficient, $\hat{b}_{j}^{(r)}(\lambda)$, of its perturbed version $X_{j}^{(r)}$.

\section{Naive GPF}
A naive implementation of the full generalized permutation framework (Naive-GPF) incorporates these permutations and is shown in Algorithm~\ref{alg:naiveGPF}.
If $T$ can be affected by stochasticity or different initial starting points of $f$ , 
then Algorithm~\ref{alg:naiveGPF} line~\ref{lst:line:permSupBaseStat} can be moved inside the $R$ loop,
so that the same random seeds or initialization are used for each $r \in R$.
\begin{algorithm}[ht]
   \caption{Permutation plug-in estimate for Naive-GPF}
   \label{alg:naiveGPF}
\belowdisplayskip=-5pt 
\begin{algorithmic}[1]
   \STATE {\bfseries Input:} response $\mathbf{y}=\{y_{i}\}_{i=1}^{N}$,  variables $\mathbf{X}= \{ x_{ij} \}_{i=1,j=1}^{N,P}$, and a supervised learning model $f$
   \STATE Sample training indices $\{\tau\}_{0 < \tau \leq N}$, with test indices being the remainder $\{1,...,N \} \setminus \{\tau\}$. \label{dataSplit}
   \STATE Split $\mathbf{y}$ and $\mathbf{X}$ according to the training and test using \ref{dataSplit}. 
   \STATE Train $f$ on the training data $\{ y_{i}, \{ x_{ij} \}_{j=1}^{P} \}_{i \in \{\tau\} }$, and compute statistic $T$ on the appropriate dataset. \label{lst:line:permSupBaseStat}
   \FOR{$\phi \in  [\Phi_{y}, \Phi_{x}] $}
     \FOR{$r=1$ {\bfseries to} $R$}   \label{lst:line:permSupBaseFunctions}
        \IF{$\phi==\Phi_{y}$ }
		    \STATE Permute the labels, $\mathbf{y'}^{(r)} =\Phi_{y}(\mathbf{y})$
		    \STATE Train $f$ on $\{ {y'}_{i}^{(r)}, \{ \mathbf{x}_{ij} \}_{j=1}^{P} \}_{i \in \{\tau\} }$, and compute the permuted data statistic $T'^{(r)}_{\Phi_{y}}$
        \ELSE
            \FOR {$p=1$ {\bfseries to} $P$}
                \STATE Permute variable $p$, $\mathbf{x'}^{(r)}_{\cdot, p} =\phi(\mathbf{x}_{\cdot, p})$, so that $\mathbf{X}^{\prime (r)}= \{ \mathbf{x}_{\cdot, 1}, ..., \mathbf{x'}_{\cdot, p}^{(r)},..., \mathbf{x}_{\cdot, P} \}$
                \STATE Train $f$ on $\{ y_{i}, \mathbf{x'}^{(r)}_{i} \}_{i \in \{\tau\} }$, and compute the permuted data statistic $T'^{(r)}_{\phi, p}$
            \ENDFOR
        \ENDIF
     \ENDFOR
   \ENDFOR
   \STATE {\bfseries Output}: test statistic distributions $T$, $\{ T'^{(r)}_{\Phi_{y}} \}_{r=1}^{R}$, $ \{ \{T'^{(r)}_{\Phi_{x}, j} \}_{r=1}^{R} \}_{j=1}^{P} $
\end{algorithmic}
\end{algorithm}

We note that the statistic $T$ can be computed on either the training data or the test data. 
If $T$ is computed on the training data, then Naive-GPF is similar to the CRT setup.
If $T$ is computed on the test data, then Naive-GPF is similar to the HRT setup.
Specifically, for linear-Gaussian and logistic synthetic data, we adopted LCD as the statistic $T$ for the training data.
For the step-wise data, we used the feature importance of the XGBoost model as the statistic $T$ on the training data.
Let $\Delta(T_{\Phi_{x}, j})$ denote the differenced null distribution for variable $j$ under $\Phi_{x}$ \eqref{eq:delta_phi_null}, 
then the one-sided p-value is given by Equation~\eqref{eq:delta_phi_pvalue}, 
where $\mathbbm{1}$ is the indicator function.
A one-sided test is used as we are only concerned with cases where the permuted data did not underperform, $T - T'^{(r)}_{\Phi_{x}, j}>0$.
\begin{align}
    \Delta(T_{\Phi_{x}, j}) &= T - \{T'^{(r)}_{\Phi_{x}, j} \}_{r=1}^{R} \label{eq:delta_phi_null} \\
    p_{j} &= \frac{1}{R+1}[1+\sum_{r=1}^{R}\mathbbm{1}(T - T'^{(r)}_{\Phi_{x}, j}>0)] \label{eq:delta_phi_pvalue}
\end{align}

\subsection{Permutation interpretations}
The first permutation $\Phi_{y}$ is the common label permutation, but applied to a supervised model $f$. 
The null hypothesis is that $f$ does not learn a statistic $T$ significantly different from the null distribution when the labels are scrambled and nothing should be learnable.
This is in the spirit of SAMseq (Algorithm~\ref{alg:permPlugin}) \citep{li2013finding}, but applied to a supervised model $f$,
whereas Algorithm~\ref{alg:permPlugin} 
has $\{f_{j}\}_{j=1}^{P}$ models, 
where each $f_{j}$ is the binary classification function applied to a variable $\mathbf{x}_{\cdot, j}$.
If the p-value from Equation~\eqref{eq:delta_phi_pvalue} for $\Delta(T_{\Phi_{y}})$ is not significant,
then model $f$ is not predictive,
and further permutation tests on variables with $\Phi_{x}$ will not be informative and are not needed.

The second permutation $\Phi_{x}$ constructs a null distribution on the effect of a variable of interest $\mathbf{x}_{\cdot, j}$ in $f$.

\subsection{Proof of subset-GPF consistency}
The output of Subset-GPF, Algorithm~\ref{alg:subsetGPF}, are the statistics $\{ \{ T^{(r)}_{j_{m}} \}_{r=1}^{R} \}_{m=1}^{M}$ and $\{ \{ T'^{(r)}_{j_{m}} \}_{r=1}^{R} \}_{m=1}^{M}$ for the variables $\{ \mathbf{x}_{\cdot, j_{m}} \}_{m=1}^{M}$ unpermuted and permuted, respectively.
Since these statistics are paired given how $X_{sub}$ are randomly sampled for each $\mathbf{x}_{\cdot,j_{m}}$ and each $r$, 
they can be re-written for a given variable $\mathbf{x}_{\cdot,j_{m}}$ as:
\begin{align}
    \{ T^{(r)}_{j_{m}} - T'^{(r)}_{j_{m}} \}_{r=1}^{R} \label{eq:diffNull_GPF2_stage1}
\end{align}
Equation~\ref{eq:diffNull_GPF2_stage1} has a similar form to the differenced null distribution for permutation $\phi$ and variable $j$ from Algorithm~\ref{alg:naiveGPF} given in Equation~\ref{eq:delta_phi_null}, which can be re-written as:  
\begin{align}
    \Delta(T_{\phi, j}) &= \{T - T'^{(r)}_{\phi, j} \}_{r=1}^{R} \label{eq:delta_phi_null2} 
\end{align}
Furthermore, if $f$ is affected by stochasticity or initial starting points, then Algorithm~\ref{alg:naiveGPF} line~\ref{lst:line:permSupBaseStat} can be moved inside the $R$ loop,
so that the same random seeds or initialization are used for each $r \in R$. 
In this case the statistic $T$ in Equation~\ref{eq:delta_phi_null2} becomes:
\begin{align}
    \Delta(T_{\phi, j}) &= \{T^{(r)} - T'^{(r)}_{\phi, j} \}_{r=1}^{R} \label{eq:delta_phi_null3} 
\end{align}
The main difference between Equations~\ref{eq:diffNull_GPF2_stage1} and \ref{eq:delta_phi_null3} is the underlying covariate matrix over which the statistic $T$ is calculated.
In Equation~\ref{eq:delta_phi_null3}, the entire covariate matrix $\mathbf{X}$, with dimension $N \times P$, is used:
\begin{align}
    T^{(r)} &= T(\mathbf{y}, f(\mathbf{X})), \\
    \text{and} \;\; T'^{(r)}_{\phi,j} &= T(\mathbf{y}, f(\mathbf{X'})), \\
    \text{where} \;\; \mathbf{X'} &= [\mathbf{x}_{\cdot,1}, ..., \phi(\mathbf{x}_{\cdot,j}), ..., \mathbf{x}_{\cdot,P}] 
\end{align}
In contrast in Equation~\ref{eq:diffNull_GPF2_stage1}, a $N \times K$ subsampled matrix $\mathbf{Z}$ is used to compute $T$ per Algorithm~\ref{alg:subsetGPF}.
Since $\mathbf{Z}$ is sampled for every $r \in R$ for variable $\mathbf{x}_{\cdot,j_{m}}$, it can be expressed more precisely as $\mathbf{Z}^{(r)}_{j_{m}} = [\mathbf{x}_{\cdot,j_{m}}, \mathbf{X}^{(r)}_{sub}]$, so that:
\begin{align}
    T^{(r)}_{j_{m}} &= T(\mathbf{y}, f(\mathbf{Z}^{(r)}_{j_{m}})), \\
    \text{and} \;\; T'^{(r)}_{j_{m}} &= T(\mathbf{y}, f(\mathbf{Z'}^{(r)}_{j_{m}})) , \\
    \text{where} \;\; \mathbf{Z'}^{(r)}_{j_{m}} &= [\Phi_{x}(\mathbf{x}_{\cdot,j_{m}}), \mathbf{X}^{(r)}_{sub}] 
\end{align}
Then $\{ \mathbf{Z}^{(r)}_{j_{m}} \}_{r=1}^{R} = \left[ \mathbf{x}_{\cdot,j_{m}}, \mathbf{X}^{(r)}_{sub} \right]_{r=1}^{R}$ are the $R$ sampled $N \times K$ matrices that include $\mathbf{x}_{\cdot,j_{m}}$, 
and $\{ T^{(r)}_{j_{m}} \}_{r=1}^{R}$ the corresponding statistics.
Let $\bar{T}_{j_{m}} = \frac{1}{R} \sum_{r=1}^{R} {T}^{(r)}_{j_{m}} $, 
then by the Law of Large Numbers:
\begin{align}
    \bar{T}_{j_{m}} &\to \mu_{j_{m}} \;\;\; \text{for} \;\;\; R \to \infty \label{eq:statMean_LLN}
\end{align}
where $\mu_{j_{m}}$ is the mean of $T$ for random $N \times K$ subsamples of $\mathbf{X}$ that include $\mathbf{x}_{\cdot,j_{m}}$.
Equation~\ref{eq:statMean_LLN} is intuitive since $\lim_{R \to \infty} \{ T^{(r)}_{j_{m}} \}_{r=1}^{R}$ is the sampling distribution of $T_{j_{m}}$.

Now consider the case for $\mathbf{x}_{\cdot,j_{m}}$ permuted, so that $\{ \mathbf{Z'}^{(r)}_{j_{m}} \}_{r=1}^{R} = \left[ \Phi_{x}(\mathbf{x}_{\cdot,j_{m}}), \mathbf{X}^{(r)}_{sub} \right]_{r=1}^{R}$.
Since there are a finite $N!$ permutations of $\mathbf{x}_{\cdot,j_{m}}=\{x_{ij}\}_{i=1}^{N}$, and the permutations are independent of the subsampling of $\mathbf{X}^{(r)}_{sub}$ from $\mathbf{X}$, the corresponding statistics $\{ T'^{(r)}_{j_{m}} \}_{r=1}^{R}$ also converges as $R \to \infty$:
\begin{align}
    \bar{T'}_{j_{m}} &= \frac{1}{R} \sum_{r=1}^{R} T'^{(r)}_{j_{m}} \\
    \bar{T'}_{j_{m}} &\to \mu'_{j_{m}} \;\;\; \text{for} \;\;\; R \to \infty \label{eq:statMean_LLN2}
\end{align}
where $\mu'_{j_{m}}$ is the mean of $T'$ for $N \times K$ subsamples of $\mathbf{X}$ that include $\Phi_{x}(\mathbf{x}_{\cdot,j_{m}})$.
This means that the sampled distribution of differenced statistics $\{ T^{(r)}_{j_{m}} - T'^{(r)}_{j_{m}} \}_{r=1}^{R}$ converges to the differenced null distribution for $K$ subsampled variables from $\mathbf{X}$ that contain $\mathbf{x}_{\cdot,j_{m}}$, 
and Subset-GPF (Algorithm~\ref{alg:subsetGPF}) yields samples from the desired null distribution.

\subsection{Calibration of subsample size $K$} 
The subsample size $K$ for Subset-GPF can be calibrated in two ways. 
First, we can balance the computational requirements of the learning model and Subset-GPF.
For example, in models, 
such as linear and generalized linear models, 
that learn a covariance structure between the covariates, 
which require $K^{2}$ observations, 
let $K=\left \lfloor{\sqrt{N} } \right \rfloor$.
Next consider $R$, the number of subsamples to run in Subset-GPF. 
The number of all subsets of size $K$ is ${P \choose K}$, and is generally too large to be computationally feasible.
A more computationally feasible choice of $R=\left( \frac{P}{K} \right) ^{2}$, 
provides an adequate mix of subset samples for each variable of interest.
This method yields the following computational cost for Subset-GPF, 
derived by taking the model training complexities in Table~\ref{table:trainingComplexity},
multiplying by the number of runs $R$, 
and substituting $R=\left( \frac{P}{K} \right) ^{2}$. 
\begin{table}[h]
\caption{Subset-GPF Complexity}
\label{table:subsetGPFComplexity}
\vskip 0.15in
\begin{center}
\begin{small}
\begin{sc}
\begin{tabular}{lcr}
\toprule
Model under Subset-GPF& $O(\cdot)$  \\
\midrule
Lasso & $O(K^{3} + N \times K^{2}) \times (\frac{P}{K})^{2} = O(K+N)P^{2}$ \\
Logistic Regression & $O(K \times C \times N \times E ) \times (\frac{P}{K})^{2} = O(\frac{P^{2}}{K}CNE)$  \\
XGBoost & $O(T \times H \times K \times N \times \log{N}) \times (\frac{P}{K})^{2} = O(\frac{P^{2}}{K}THN\log{N}) $ \\
\bottomrule
\end{tabular}
\end{sc}
\end{small}
\end{center}
\vskip -0.1in
\end{table}


The second way to choose the subsample size $K$ is given by the desired goal of Subset-GPF (Algorithm~\ref{alg:subsetGPF}), 
which is to identify variables that significantly affect model predictive performance,
by relating the subsample size to the model performance metric $T$.
This is done through Algorithm~\ref{alg:K_calibration}.
We can use a range around $K=\left \lfloor{\sqrt{N} } \right \rfloor$ for $K_{0}$ and $K_{1}$.
\begin{algorithm}
   \caption{Calibrating subsample size $K$}
   \label{alg:K_calibration}
\belowdisplayskip=-5pt 
\begin{algorithmic}[1]
   \STATE {\bfseries Input:} labels $\mathbf{y}=\{y_{i}\}_{i=1}^{N}$, covariates $\mathbf{X}= \{ x_{ij} \}_{i=1, j=1}^{N, P}$, and supervised model $f$
   \FOR{$k=K_{0}$ {\bfseries to} $K_{1}$}
       \FOR{$r=1$ {\bfseries to} $R$}
            \STATE Subsample $\mathbf{Z}^{(r)}_{k}$, dimension $N \times K$ from $\mathbf{X}$ randomly.
            \STATE Split $\mathbf{y}$ and $\mathbf{Z}^{(r)}_{k}$ into training and test sets. \label{lst:line:kcalibration_train_test_split}
            \STATE Train and compute the test statistic $T^{(r)}_{k}$ from $\mathbf{y}$,  $\mathbf{Z}^{(r)}_{k}$ and $f$.
    		\STATE Permute the labels $\mathbf{y'}^{(r)} = \Phi_{y}(\mathbf{y})$
    		\STATE Train and compute the test statistic $T'^{(r)}_{\Phi_{y}, k}$ from $\mathbf{y}'^{(r)}$,  $\mathbf{Z}^{(r)}_{k}$ and $f$.
    \ENDFOR
   \ENDFOR
   \STATE {\bfseries Output}: $\{ \{T'^{(r)}_{\Phi_{y}, k} \}_{r=1}^{R} \}_{k=K_{0}}^{K_{1}}$ , the null distribution of $\{ \{T^{(r)}_{k} \}_{r=1}^{R} \}_{k=K_{0}}^{K_{1}}$ 
\end{algorithmic}
\end{algorithm}
The intuition is to find a $K_{0} \leq k \leq K_{1}$, where model $f$ is significantly predictive of $T$ on subsamples $\mathbf{Z}$ compared to the label scrambled null distribution \eqref{eq:delta_phi_pvalue_label}.
\begin{align}
    \Delta(T_{\Phi_{y}, k}) &= \{T^{(r)}_{k} - T'^{(r)}_{\Phi_{y}, k} \}_{r=1}^{R} \label{eq:delta_phi_null_label} \\
    \text{p-value} &= \frac{1}{R}\sum_{r=1}^{R}I(T^{(r)}_{k} - T'^{(r)}_{\Phi_{y},k}<0) \label{eq:delta_phi_pvalue_label}
\end{align}
Note the upper limit for $K_{1}$ is $K_{1}= P$.

\subsection{Calibrating the number of perturbations $R$ in subset-GPF for a desired TPR level}
We show how to calibrate the TPR or power of subset-GPF by choosing $R$ for a given $K$ for two major types of response and variable relationships.
First, we consider linear-Gaussian relationships:
\begin{align}
	\mathbf{X} &\sim \mathcal{N}(0, \text{diag}({\sigma_{x}^{2}}))  \\
	y_{i} &= \sum_{j=1}^{P} \beta_{j} x_{ij} + \epsilon_{i}, \quad \text{where } \epsilon_{i} \sim \mathcal{N}(0, \sigma^{2}_{\mathbf{y}}) \\
	\mathbf{y} &\sim \mathcal{N}(\beta \mathbf{X}, \sigma^{2}_{\mathbf{y}}) 
\end{align}
For this setup, the coefficients for the linear model $f$ are estimated as:
\begin{align}
	\hat{\beta} &= (\mathbf{X}^{\top}\mathbf{X})^{-1} \mathbf{X}^{\top} \mathbf{y} \\
	\beta & \sim \mathcal{N}(\hat{\beta}, (\mathbf{X}^{\top}\mathbf{X})^{-1} \sigma^{2}_{\mathbf{y}}) 
\end{align}
The coefficient for the unperturbed variable $x_{j}$ is $\hat{\beta_{j}}$.
In the full naive-GPF, the variable $x_{j}$ is permuted to become $x_{j}^{(r)} = \Phi_{x}(x_{j})$, and the estimate of the coefficient becomes:
\begin{align}
	\mathbf{X}^{\prime (r)} &= \{ \mathbf{x}_{\cdot, 1}, ..., \mathbf{x}_{\cdot, j}^{\prime (r)},..., \mathbf{x}_{\cdot, P} \} \\
	\hat{\beta}^{\prime} &= (\mathbf{X}^{\prime (r) \top}\mathbf{X}^{\prime (r)})^{-1} \mathbf{X}^{\prime (r) \top} \mathbf{y} 
\end{align}
Then the coefficient of the perturbed variable $x_{j}^{\prime (r) }$ is $\hat{\beta_{j}^{\prime}}$.
One nuance is that the distribution of the $\beta^{\prime}$ depends on whether $x_{j}$ is in the set of variables of real explanatory variables $S$:
\begin{equation}	
	\beta^{\prime} \sim 
	\begin{cases} 
		\mathcal{N}(\hat{\beta}, (\mathbf{X}^{\top}\mathbf{X})^{-1} \sigma^{2}_{\mathbf{y}})  & x_{j} \notin S \\
		\mathcal{N}(\hat{\beta^{\prime }}, ( (\mathbf{X}^{\top}\mathbf{X})^{-1} \sigma^{2}_{\mathbf{y}} + (\beta_{j}x_{j})^{2} )  & x_{j} \in S
	\end{cases}	
\end{equation}
If $x_{j} \in S$ and $x_{j}$ is perturbed, then $\mathbf{y} - \beta \mathbf{X}^{\prime} = \mathbf{\epsilon} + \beta_{j}x_{j}$.
That is the residual will be bigger by $\beta_{j}x_{j}$ even if the true linear weights $\beta$ are used.
This result can be easily verified by simulation.

The full GPF uses the LCD to test the significance of variable $x_{j}$: $\text{LCD}_{j}^{(r)} = |\hat{\beta}_{j}| - |\hat{\beta_{j}^{\prime}}|$.
Then for the LCD to have 90\% TPR at $\alpha=0.05$ from a 1-sided t-distribution, the number of perturbation simulations for variable $x_{j}$, assuming that $x_{j}$ is in the set of explanatory variables $S$ is given by the rule of thumb \citep{lehr1992sixteen, van2011statistical}:
\begin{align}
	R &=11 \times \frac{\text{var}(\hat{\beta^{\prime}})} {(\mathbb{E}(\text{LCD}))^{2}}  \\ 
	\mathbb{E}_{r}(\text{LCD}_{j}) &= \mathbb{E}_{r}(|\hat{\beta_{j}}| - |\hat{\beta}_{j}^{\prime}|)  =  |\hat{\beta_{j}}| = \beta_{j} \\
	\text{var}(\hat{\beta^{\prime}}) &=  (\mathbf{X}^{\top}\mathbf{X})^{-1} \sigma^{2}_{y} + (\beta_{j}x_{j})^{2} \\
				& \approx (N^{2} \sigma_{x}^{2})^{-1} \sigma_{y}^{2} + (\beta_{j} \sigma_{x})^{2} \\	
	\implies R  & \approx 11 \times \frac{(\beta_{j} \sigma_{x})^{2}}{\beta_{j}^{2}}
\end{align}

Now consider the case of subset-GPF, which subsamples $K-1$ variables uniform randomly and combines them with a variable of interest $x_{j}$.
The subset coefficient $\beta_{sub}$ is given by:
\begin{align}
	\hat{\beta}_{sub} &= (\mathbf{X}_{sub}^{\top}\mathbf{X}_{sub})^{-1} \mathbf{X}_{sub}^{\top} \mathbf{y} \\
	\beta_{sub} & \sim \mathcal{N}\left( \hat{\beta}_{sub}, \left( (\mathbf{X}_{sub}^{\top}\mathbf{X}_{sub})^{-1} \sigma^{2}_{\mathbf{y}} + \left( \frac{P-K+1}{P-1} \times |S| \bar{\beta} \right)^{2} \right) \right) 
\end{align}
The $\frac{P-K+1}{P-1} \times |S| \bar{\beta}$ adjustment accounts for how many explanatory variables from $S$ would not be included in uniformly randomly chosen subsamples of size $K-1$ from the set of $P-1$ variables not including $x_{j}$, with $\bar{\beta}$ the average weight of the explanatory variables.
Next, if $x_{j}$ is perturbed, the distribution of $\beta^{\prime}_{sub}$ is:
\begin{align}
	\hat{\beta}^{\prime}_{sub} &= (\mathbf{X}^{\prime (r) \top}_{sub}\mathbf{X}^{\prime (r)}_{sub})^{-1} \mathbf{X}^{\prime (r) \top}_{sub} \mathbf{y} 
\end{align}
and the distribution of $\beta^{\prime}_{sub}$ is:
\begin{equation}	
	\beta^{\prime}_{sub} \sim 
	\begin{cases} 
		\mathcal{N}\left( \hat{\beta}_{sub}, \left( (\mathbf{X}_{sub}^{\top}\mathbf{X}_{sub})^{-1} \sigma^{2}_{\mathbf{y}} + \left( \frac{P-K+1}{P-1} \times |S| \bar{\beta} \right)^{2} \right) \right)  \notin S \\
		\mathcal{N}\left( \hat{\beta}^{\prime}_{sub}, \left( (\mathbf{X}_{sub}^{\top}\mathbf{X}_{sub})^{-1} \sigma^{2}_{\mathbf{y}} + \left( \frac{P-K+1}{P-1} \times |S| \bar{\beta} \right)^{2} + (\beta_{j}x_{j})^{2} \right) \right)   & x_{j} \in S
	\end{cases}	
\end{equation}
Then the subset-GPF LCD for $x_{j}$ is $|\hat{\beta}_{j,sub}| - |\hat{\beta}^{\prime}_{j,sub}|$.
For this LCD to be powered at the 90\% level or have 90\% TPR with $\alpha=0.05$, the number of runs $R$, based on the 1-sided t-distribution, should be: $R=11 \times \frac{\text{var}(\hat{\beta}_{sub}^{\prime}) }{\mathbb{E}(\text{LCD})^{2}} \approx 11 \left( \frac{P-K}{P} \times |S| \bar{\beta} \right)^{2} / \beta_{j}^{2}$.
For our experiments on synthetic linear-Gaussian and logistic data, with $N=250$, $P=400$, $|S|=20$, common $\beta$ for variables in the explanatory variable set $S$, so that $\bar{\beta} = \beta_{j}$, and $K=\left \lfloor{\sqrt{N} } \right \rfloor = 15$, $R \approx 3970$ for 90\% TPR.
More generally, additive models that are not linear in $x_{j}$ can be approximated by linear models by adding transformations of $x_{j}$ to the design matrix $\mathbf{X}$.
These added transformations will increase the variable dimension $P$ to $P^{\prime}$, and the $R$ calibration calculation should be adjusted for $P^{\prime}$.

Next, consider the class of models that are not additive, but links the response $\mathbf{y}$ to a non-linear function of at least two explanatory variables.
Suppose that when two or more explanatory variables included in training a model, the model learns and predicts held-out test responses $\mathbf{t}_{\text{test}}$ with much lower MSE for real-valued $\mathbf{y}$ or F1 score for categorical $\mathbf{y}$.
Then in subset-GPF, we require that random samples of $K-1$ variables contain at least one other explanatory variable with 90\% probability to complement the variable being studied through permutation perturbations to detect significant differences in model performance.
Therefore $K$ can be chosen based on $S$ or $\hat{S}$ an estimate if $S$ unknown so that $K$ satisfies: $90\% = 1-\left( \frac{P-S}{P} \right)^{K-1}$.
This amounts to the shared birthday problem,
and shows how we can effectively cover the space $P$ variables with $R$ $P \choose K$ subsamples.

\section{SAMseq algorithm}
\begin{algorithm}
   \caption{Permutation plug-in estimate}
   \label{alg:permPlugin}
\belowdisplayskip=-5pt 
\begin{algorithmic}[1]
   \STATE {\bfseries Input:} labels $\mathbf{y}=\{y_{i}\}_{i=1}^{N}$, covariates $\mathbf{X}= \{ x_{ij} \}_{i=1, j=1}^{N, P}$. Let variable $\mathbf{x}_{\cdot, j} = \{x_{ij}\}_{i=1}^{N}$
   \STATE Compute the test statistics $\{T_{j} \}_{j=1}^{P}$ from $\mathbf{y}$ for each variable $\mathbf{x}_{\cdot, j}$
   \FOR{$r=1$ {\bfseries to} $R$}
		\STATE Permute the labels $\mathbf{y}$ to get $\mathbf{y'}^{(r)}$
		\STATE Compute $\{{T'}_{j}^{(r)} \}_{j=1}^{P}$ from the permuted labels  $\mathbf{y'}^{(r)}$ and unchanged variables $\{ \mathbf{x}_{\cdot, j} \}_{j=1}^{P}$
   \ENDFOR
   \STATE {\bfseries Output}: $\{ \{{T'}_{j}^{(r)} \}_{r=1}^{R} \}_{j=1}^{P}$, the null distribution of $\{T_{j} \}_{j=1}^{P}$ 
\end{algorithmic}
\end{algorithm}
\citet{li2013finding} considered the case of $N$ observations of $\{y_{i}\}_{i=1,...,N}$ belonging to one of two classes, $C_{1}$ or $C_{2}$ and $N$ observations of $P$ variables $\{x_{ij}\}_{i=1,...,N; j=1,...,P}$. The ranks of each variable $x_{j}$ and the two-sample Wilcoxon statistic is used as $T_{j}^{k}$ for each class:
\begin{align}
	R(X_{ij}) &= \text{rank of } x_{ij} \text{ in } \{x_{ij}\}_{i=1,...,N} \\
	T_{j}^{k} &= \sum_{y_{i} \in C_{k}} R(X_{ij}) - \frac{|C_{k}| (N+1)}{2}
\end{align}
Note that the term $\frac{|C_{k}|(N+1)}{2}$ differs from the standard Wilcoxon rank statistic as it adjusts for class-imbalanced datasets through the class size $|C_{k}|$.
The corresponding statistic for the $r^{th}$ label permutated data is:
\begin{align}
	T_{j}^{\prime k(r)} &= \sum_{y_{i}^{\prime (r)} \in C_{k}} R(X_{ij}) - \frac{|C_{k}| (N+1)}{2}
\end{align}
The p-value of each $x_{j}$ is computed by comparing $T_{j}^{k}$ to $T_{j}^{\prime k (r)}$:
\begin{align}
	p_{j}^{k} &= \frac{1}{R+1} \left[ 1+\sum_{r=1}^{R} \mathbbm{1}(|T_{j}^{\prime k(r)}| \geq |T_{j}^{k}|) \right]
\end{align}
SAMseq estimates the probability a variable $x_{j}$ is over or under-expressed for a class $k$ by considering how likely the distribution of its ranksum would occur given random uniform permutations of the $y$ class labels.
As SAMseq only considers ranks of observed variables, it makes no distributional assumptions and is non-parametric.
However, SAMseq can only detect over or under-expressed variables.
It does not consider more complicated relationships between $\mathbf{y}$ and $\mathbf{X}$ through a model $f$.
We note SAMseq could have been implemented by permuting the variables $x_{j}$ while keeping $\mathbf{y}$ unchanged and computing ranks on the permuted $x_{j}$.
It was implemented by permuting $\mathbf{y}$ because that breaks the relationship between the labels $\mathbf{y}$ and all the variables $\mathbf{X}$ and is more computationally efficient.
However, if a more complicated functional model $f$ is learnt between $\mathbf{y}$ and $\mathbf{X}$, then one would want to ascertain the contribution of individual variables $x_{j}$ to learning $f$ given $\mathbf{y}$ and all the other variables $\{x_{i}\}_{i \ne j}$.
In this case, the $x_{j}$ need to be permuted separately while the labels are kept unchanged. 
This is the approach adopted by GPF, CRT and HRT.

\begin{figure}[htbp]
    \centering
    \includegraphics[scale=.59]{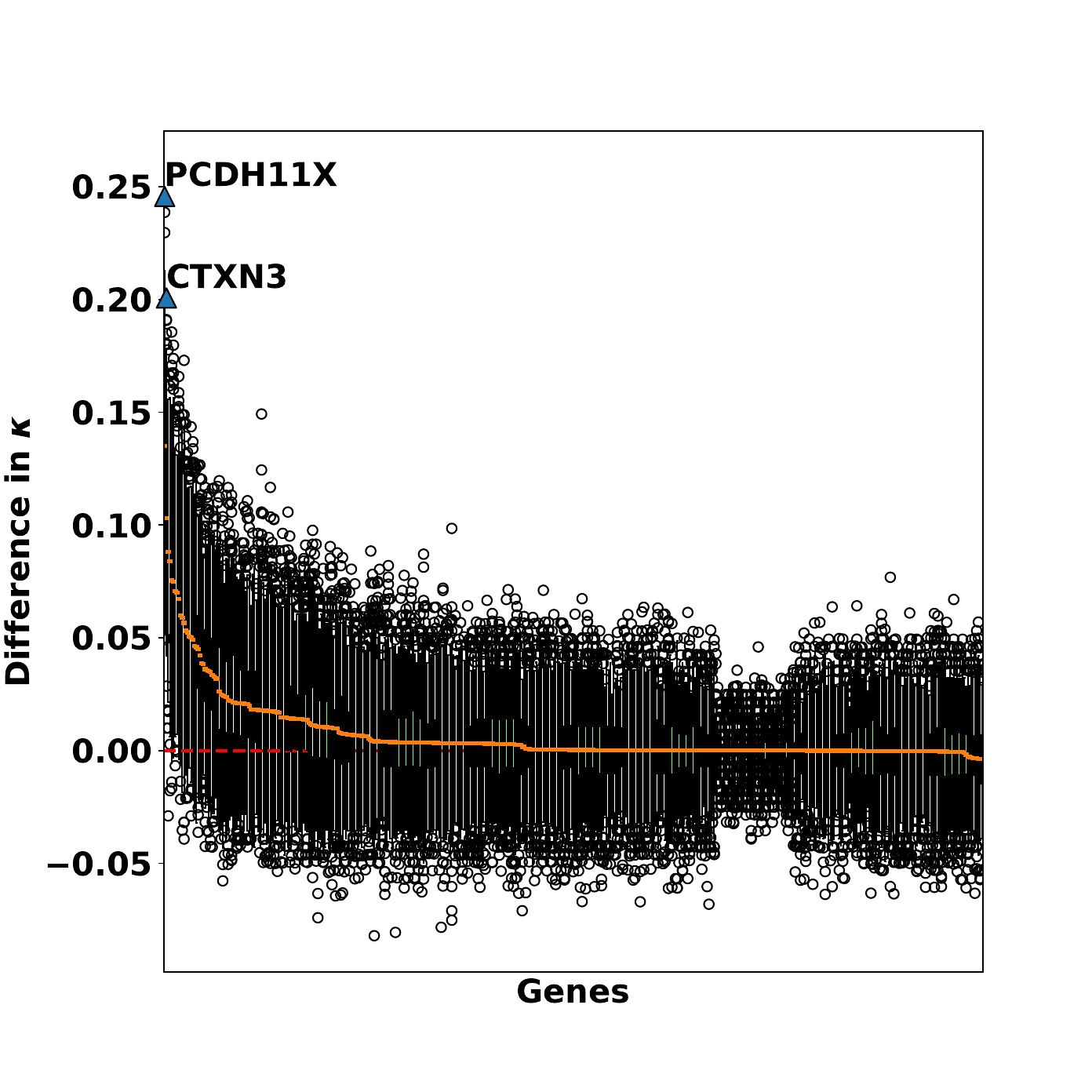}
    \caption{$\Delta(\kappa)$ for the $500$ genes. Green shaded boxes indicate randomly selected genes. Significant genes at $\alpha=\frac{0.05}{50000}$ are shaded blue and labeled.}
    \label{fig:allen_paired_kappa_boxplot_all}
\end{figure}
\begin{figure}[htbp]
    \centering
    \includegraphics[scale=0.75]{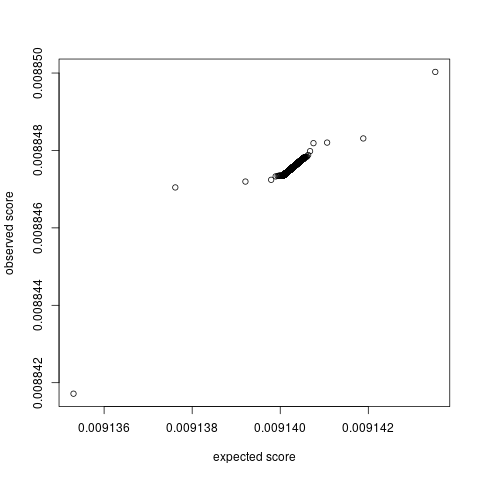}
    \caption{Q-Q plot for SAMseq for multiclass prediction of brain region using the Allen brain gene-expression dataset. Of note, none of the genes are significantly different from the null distribution.}
    \label{fig:allen_samseq}
\end{figure}

\end{document}